\documentclass[letterpaper,12pt]{JHEP3}
\usepackage{amsmath,amssymb}
\usepackage{latexsym}
\usepackage{epsfig}
\usepackage{cite}
\usepackage[english]{babel}

\newcommand{\X}{X\!\!\cdot\!\xi}
\newcommand{\Xb}{{\bar X}\!\!\cdot\!\xi}

\newcommand{\eq}{\begin{equation}}
\newcommand{\feq}{\end{equation}}
\newcommand{\eqn}{\begin{eqnarray}}
\newcommand{\feqn}{\end{eqnarray}}
\newcommand{\arr}{\begin{eqnarray*}}
\newcommand{\farr}{\end{eqnarray*}}

\font\mybb=msbm10 at 12pt
\def\bb#1{\hbox{\mybb#1}}
\def\bZ {\bb{Z}}
\def\bI {\bb{I}}
\def\bR {\bb{R}}

\def\bC {\bb{C}}

\title{All null supersymmetric backgrounds of ${\cal N}=2$, $D=4$ gauged
supergravity coupled to abelian vector multiplets}

\author{Dietmar Klemm and Emanuele Zorzan \\
Dipartimento di Fisica dell'Universit\`a di Milano, \\
\hspace*{0.15cm} Via Celoria 16, I-20133 Milano and \\
INFN, Sezione di Milano, Via Celoria 16, I-20133 Milano. \\
}
\preprint{IFUM-933-FT}

\abstract{The lightlike supersymmetric solutions of ${\cal N}=2$, $D=4$ gauged
supergravity coupled to an arbitrary number of abelian vector multiplets are
classified using spinorial geometry techniques. The solutions fall into two
classes, depending on whether the Killing spinor is constant or not. In both
cases, we give explicit examples of supersymmetric backgrounds.
Among these BPS solutions, which preserve one quarter of the supersymmetry,
there are gravitational waves propagating on domain
walls or on bubbles of nothing that asymptote to AdS$_4$. Furthermore, we
obtain the additional constraints obeyed by half-supersymmetric vacua.
These are divided into four categories, that include bubbles of nothing
which are asymptotically AdS$_4$, pp-waves on domain walls, AdS$_3$ $\times$
$\bR$, and spacetimes conformal to AdS$_3$ times an interval.
}


\begin{document}

\section{Introduction}
\label{intro}

Supersymmetric solutions to supergravity theories have played, and continue to
play, an important role in string- and M-theory developments. This makes it
desirable to obtain a complete classification of BPS solutions to various
supergravities in diverse dimensions. Progress in this direction has
been made in the last years using the mathematical concept of
G-structures \cite{Gauntlett:2002sc}.
The basic strategy is to assume the existence of at least one Killing spinor
$\epsilon$ obeying ${\cal D}_{\mu}\epsilon=0$, and to
construct differential forms as bilinears from this spinor. These forms, which
define a preferred G-structure, obey several algebraic and differential equations
that can be used to deduce the metric and the other bosonic supergravity fields.
Using this framework, a number of complete
classifications \cite{Gauntlett:2002nw, Gutowski:2003rg, Meessen:2006tu} and many
partial results (see e.g.~\cite{Gauntlett:2002fz, Gauntlett:2003fk, Caldarelli:2003pb, Caldarelli:2003wh, Gauntlett:2003wb, Cariglia:2004kk, Cacciatori:2004rt, Cariglia:2004qi,
Gutowski:2005id, Bellorin:2005zc, Huebscher:2006mr, Bellorin:2006yr, Bellorin:2007yp}
for an incomplete list) have been obtained. By complete we mean that the most
general solutions for all possible fractions of supersymmetry have been obtained,
while for partial classifications this is only available for some
fractions. Note that the complete classifications mentioned above involve
theories with eight supercharges and holonomy $H=$ SL(2,$\mathbb{H}$) of the
supercurvature $R_{\mu\nu} = {\cal D}_{[\mu} {\cal D}_{\nu ]}$, and allow for either
half- or maximally supersymmetric solutions.

An approach which exploits the linearity of the Killing spinors has
been proposed \cite{Gillard:2004xq} under the name of spinorial
geometry. Its basic ingredients are an explicit oscillator basis for
the spinors in terms of forms and the use of the gauge symmetry to
transform them to a preferred representative of their orbit.
While the equivalent G-structure technique leads to nonlinear equations which
might be difficult to interpret and to solve in some cases, the spinorial
geometry approach permits to construct a linear system for the background
fields from any (set of) Killing spinor(s) \cite{Gran:2005wu}. This
method has proven fruitful in e.g.~the challenging case of IIB
supergravity \cite{Gran:2005wn, Gran:2005kg, Gran:2005ct}. In
addition, it has been adjusted to impose 'near-maximal'
supersymmetry and thus has been used to rule out certain large fractions of
supersymmetry \cite{Gran:2006ec, Grover:2006ps, Grover:2006wy,
Gran:2006cn, Gran:2007eu}. Finally, a complete classification for type I
supergravity in ten dimensions has been obtained in \cite{Gran:2007fu}, and all
half-supersymmetric backgrounds of ${\cal N}=2$, $D=5$ gauged supergravity
coupled to abelian vector multiplets were determined
in \cite{Gutowski:2007ai, Grover:2008ih}. Spinorial geometry was also
applied to de~Sitter supergravity \cite{Grover:2008jr}, where interesting
mathematical structures like hyper-K\"ahler manifolds with torsion emerge.

In the present paper we shall finish the classification of
supersymmetric solutions in four-dimensional ${\cal N}=2$ matter-coupled
U(1)-gauged supergravity initiated in \cite{Cacciatori:2008ek}, generalizing
thus the simpler cases of ${\cal N}=1$, considered recently
in \cite{Ortin:2008wj,Gran:2008vx}, and minimal ${\cal N}=2$, where a full
classification is available both in the ungauged \cite{Tod:1983pm}
and gauged theories \cite{Cacciatori:2007vn}. A strong motivation for our
work comes from the AdS$_4$/CFT$_3$ correspondence, which has been attracting
much attention in the last months, after
the discovery of superconformal field theories describing coincident
M2-branes \cite{Bagger:2007jr,Aharony:2008ug}. In this context, supergravity
vacua with less supersymmetry correspond on the CFT side to nonzero vacuum
expectation values of certain operators, or to deformations of the CFT.
Disposing of a systematic classification of supergravity vacua is thus particularly
useful. Of special interest in this context are domain wall solutions
interpolating between vacua preserving different amounts of
supersymmetry, because they can describe a holographic RG flow.

The case where the Killing vector constructed from the Killing spinor is
timelike was considered in \cite{Cacciatori:2008ek}, so we will now concentrate
on the null class. Note that this is more than a mere extension of
\cite{Cacciatori:2008ek}: The timelike case typically contains black hole
solutions, while the lightlike class includes gravitational waves and domain walls,
whose importance in an AdS/CFT context was just explained.

The outline of this paper is as follows. In section \ref{sugra}, we briefly
review ${\cal N}=2$ supergravity in four dimensions and its matter couplings,
whereas in \ref{orbits} the orbits of Killing spinors are discussed.
In section \ref{null} we determine the conditions
coming from a single null Killing spinor and give explicit examples for
supersymmetric backgrounds. Finally, in section \ref{half-susy}, we impose
a second Killing spinor and obtain the additional constraints obeyed by
half-supersymmetric solutions. It is shown that half-BPS geometries are
divided into four classes, that include bubbles of nothing
which are asymptotically AdS$_4$, pp-waves on domain walls, AdS$_3$ $\times$
$\bR$, and spacetimes conformal to AdS$_3$ times an interval.
Appendices \ref{conv} and \ref{spin-forms} contain
our notation and conventions for spinors.

\section{Matter-coupled ${\cal N}=2$, $D=4$ gauged supergravity}
\label{sugra}

In this section we shall give a short summary
of the main ingredients of ${\cal N}=2$, $D=4$ gauged supergravity coupled
to vector- and hypermultiplets \cite{Andrianopoli:1996cm}. Throughout this paper,
we will use the notations and conventions of \cite{Vambroes}, to which we refer
for more details.

Apart from the vierbein $e^a_{\mu}$ and the chiral gravitinos $\psi^i_{\mu}$,
$i=1,2$, the field content includes $n_H$ hypermultiplets and $n_V$ vector multiplets
enumerated by $I=0,\ldots,n_V$. The latter contain the graviphoton and have
fundamental vectors $A^I_{\mu}$, with field strengths
\begin{displaymath}
F^I_{\mu\nu} = \partial_{\mu}A^I_{\nu} - \partial_{\nu}A^I_{\mu} + gA^K_{\nu}A^J_{\mu}
              {f_{JK}}^I\,.
\end{displaymath}
The fermions of the vector multiplets are denoted as $\lambda^{\alpha i}$ and the
complex scalars as $z^{\alpha}$ where $\alpha=1,\ldots,n_V$. These scalars parametrize
a special K\"ahler manifold, i.~e.~, an $n_V$-dimensional
Hodge-K\"ahler manifold that is the base of a symplectic bundle, with the
covariantly holomorphic sections
\begin{equation}
{\cal V} = \left(\begin{array}{c} X^I \\ F_I\end{array}\right)\,, \qquad
{\cal D}_{\bar\alpha}{\cal V} = \partial_{\bar\alpha}{\cal V}-\frac 12
(\partial_{\bar\alpha}{\cal K}){\cal V}=0\,, \label{sympl-vec}
\end{equation}
where ${\cal K}$ is the K\"ahler potential and ${\cal D}$ denotes the
K\"ahler-covariant derivative\footnote{For a generic field $\phi^{\alpha}$ that
transforms under a K\"ahler transformation ${\cal K}(z,\bar z)\to
{\cal K}(z,\bar z)+\Lambda(z)+\bar\Lambda(\bar z)$ as
$\phi^{\alpha}\to e^{-(p\Lambda+q\bar\Lambda)/2}\phi^{\alpha}$, one has
${\cal D}_{\alpha}\phi^{\beta}=\partial_{\alpha}\phi^{\beta}+
{\Gamma^{\beta}}_{\alpha\gamma}\phi^{\gamma}+\frac p2(\partial_{\alpha}{\cal K})
\phi^{\beta}$. ${\cal D}_{\bar\alpha}$ is defined in the same
way. $X^I$ transforms as $X^I\to e^{-(\Lambda -\bar\Lambda)/2}X^I$ and
thus has K\"ahler weights $(p,q)=(1,-1)$.}.
${\cal V}$ obeys the symplectic constraint
\begin{equation}
\langle {\cal V}\,,\bar{\cal V}\rangle = X^I\bar F_I-F_I\bar X^I=i\,.
\end{equation}
To solve this condition, one defines
\begin{equation}
{\cal V}=e^{{\cal K}(z,\bar z)/2}v(z)\,,
\end{equation}
where $v(z)$ is a holomorphic symplectic vector,
\begin{equation}
v(z) = \left(\begin{array}{c} Z^I(z) \\ \frac{\partial}{\partial Z^I}F(Z)
\end{array}\right)\,.
\end{equation}
F is a homogeneous function of degree two, called the prepotential,
whose existence is assumed to obtain the last expression.
This is not restrictive because it can be shown
that it is always possible to go in a gauge where the prepotential exists via a local symplectic
transformation \cite{Vambroes,Craps:1997gp}\footnote{This need not
be true for gauged supergravity, where symplectic covariance is broken
\cite{Andrianopoli:1996cm}. However, in our analysis we do not really use
that the $F_I$ can be obtained from a prepotential, so
our conclusions go through also without assuming that $F_I=\partial
F(X)/\partial X^I$ for some $F(X)$.
We would like to thank Patrick Meessen for discussions on this point.}.
The K\"ahler potential is then
\begin{equation}
e^{-{\cal K}(z,\bar z)} = -i\langle v\,,\bar v\rangle\,.
\end{equation}
The matrix ${\cal N}_{IJ}$ determining the coupling between the scalars $z^{\alpha}$
and the vectors $A^I_{\mu}$ is defined by the relations
\begin{equation}\label{defN}
F_I = {\cal N}_{IJ}X^J\,, \qquad {\cal D}_{\bar\alpha}\bar F_I = {\cal N}_{IJ}
{\cal D}_{\bar\alpha}\bar X^J\,.
\end{equation}
Given
\begin{equation}
U_{\alpha} \equiv {\cal D}_{\alpha}{\cal V} = \partial_{\alpha}{\cal V} + \frac 12
(\partial_{\alpha}{\cal K}){\cal V}\,,
\end{equation}
the following differential constraints hold:
\begin{eqnarray}
{\cal D}_{\alpha}U_{\beta} &=& C_{\alpha\beta\gamma}g^{\gamma\bar\delta}\bar U_{\bar\delta}\,,
\nonumber \\
{\cal D}_{\bar\beta}U_{\alpha} &=& g_{\alpha\bar\beta}{\cal V}\,, \nonumber \\
\langle U_{\alpha}\,, {\cal V}\rangle &=& 0\,.
\end{eqnarray}
Here, $C_{\alpha\beta\gamma}$ is a completely symmetric tensor which determines also
the curvature of the special K\"ahler manifold.

We now come to the hypermultiplets. These contain scalars $q^X$ and spinors
$\zeta^A$, where $X=1,\ldots,4n_H$ and $A=1,\ldots,2n_H$. The $4n_H$ hyperscalars
parametrize a quaternionic K\"ahler manifold, with vielbein $f^{iA}_X$ and inverse
$f^X_{iA}$ (i.~e.~the tangent space is labelled by indices $(iA)$). From these one
can construct the three complex structures
\begin{equation}
\vec J_X^{\;\;\;Y} = -if^{iA}_X\vec\sigma_i^{\;\;j} f^Y_{jA}\,,
\end{equation}
with the Pauli matrices $\vec\sigma_i^{\;\;j}$ (cf.~appendix \ref{conv}).
Furthermore, one defines SU$(2)$ connections $\vec\omega_X$ by requiring the
covariant constancy of the complex structures:
\begin{equation}
0 = {\mathfrak D}_X\vec J_Y^{\;\;\;Z}\equiv \partial_X\vec J_Y^{\;\;\;Z}-
{\Gamma^W}_{XY}\vec J_W^{\;\;\;Z}+{\Gamma^Z}_{XW}\vec J_Y^{\;\;\;W}+
2\,\vec\omega_X\times\vec J_Y^{\;\;\;Z}\,,
\end{equation}
where the Levi-Civita connection of the metric $g_{XY}$ is used. The curvature
of this SU$(2)$ connection is related to the complex structure by
\begin{equation}
{\vec R}_{XY} \equiv 2\,\partial_{\left[X\right.}\vec\omega_{\left.Y\right]} +
2\,\vec\omega_X\times\vec\omega_Y = -\frac 12\kappa^2{\vec J}_{XY}\,.
\end{equation}
Depending on whether $\kappa=0$ or $\kappa\neq 0$ the manifold is hyper-K\"ahler
or quaternionic K\"ahler respectively. In what follows, we take $\kappa=1$.

The bosonic action of ${\cal N}=2$, $D=4$ supergravity is
\begin{eqnarray}
e^{-1}{\cal L}_{\text{bos}} &=& \frac 1{16\pi G}R + \frac 14(\text{Im}\,{\cal N})_{IJ}
F^I_{\mu\nu}F^{J\mu\nu} - \frac 18(\text{Re}\,{\cal N})_{IJ}\,e^{-1}
\epsilon^{\mu\nu\rho\sigma}F^I_{\mu\nu}F^J_{\rho\sigma}\,, \nonumber \\
&& -g_{\alpha\bar\beta}{\cal D}_{\mu}z^{\alpha}{\cal D}^{\mu}\bar z^{\bar\beta} -
\frac 12 g_{XY}{\cal D}_{\mu}q^X{\cal D}^{\mu}q^Y - V\,, \nonumber \\
&& -\frac g6 C_{I,JK}e^{-1}\epsilon^{\mu\nu\rho\sigma}A^I_{\mu}A^J_{\nu}
(\partial_{\rho}A^K_{\sigma}-\frac 38 g{f_{LM}}^KA^L_{\rho}A^M_{\sigma})\,,
\label{action}
\end{eqnarray}
where $C_{I,JK}$ are real coefficients, symmetric in the last two indices,
with $Z^I Z^J Z^K C_{I,JK}=0$, and the covariant derivatives acting on the
scalars read
\begin{equation}
{\cal D}_{\mu}z^{\alpha} = \partial_{\mu}z^{\alpha} + g A^I_{\mu}k^{\alpha}_I(z)\,,
\qquad {\cal D}_{\mu}q^X = \partial_{\mu}q^X + g A^I_{\mu}k^X_I\,.
\end{equation}
Here $k^{\alpha}_I(z)$ and $k^X_I(q)$ are Killing vectors of the special K\"ahler
and quaternionic K\"ahler manifolds respectively. The potential $V$
in \eqref{action} is the sum of three distinct contributions:
\begin{eqnarray}
V &=& g^2(V_1+V_2+V_3)\,, \nonumber \\
V_1 &=& g_{\alpha\bar\beta}k^{\alpha}_I k^{\bar\beta}_J\,e^{\cal K}\bar Z^I Z^J\,,
\nonumber \\
V_2 &=& 2\,g_{XY}k^X_I k^Y_Je^{\cal K}\bar Z^I Z^J\,, \nonumber \\
V_3 &=& 4(U^{IJ}-3\,e^{\cal K}\bar Z^I Z^J)\vec P_I\cdot\vec P_J\,, \label{scal-pot}
\end{eqnarray}
with
\begin{equation}
U^{IJ} \equiv g^{\alpha\bar\beta}e^{\cal K}{\cal D}_{\alpha}Z^I{\cal D}_{\bar\beta}
\bar Z^J = -\frac 12(\text{Im}\,{\cal N})^{-1|IJ}-e^{\cal K}\bar Z^I Z^J\,,
\label{invImN}
\end{equation}
and the triple moment maps $\vec P_I(q)$. The latter have to satisfy the
equivariance condition
\begin{equation}
\vec P_I\times\vec P_J + \frac 12 \vec J_{XY}k^X_Ik^Y_J - {f_{IJ}}^K
\vec P_K = 0\,, \label{equivariance}
\end{equation}
which is implied by the algebra of symmetries. The metric for the vectors
is given by
\begin{equation}
{\cal N}_{IJ}(z,\bar z) = \bar F_{IJ} + i\frac{N_{IN}N_{JK}Z^NZ^K}{N_{LM}Z^LZ^M}\,,
\qquad N_{IJ} \equiv 2\,\text{Im}\,F_{IJ}\,,
\end{equation}
where $F_{IJ}=\partial_I\partial_JF$, and $F$ denotes the prepotential.

Finally, the supersymmetry transformations of the fermions to bosons are
\begin{eqnarray}
\delta\psi^i_{\mu} &=& D_{\mu}(\omega)\epsilon^i - g\Gamma_{\mu}S^{ij}\epsilon_j
+ \frac 14\Gamma^{ab}F^{-I}_{ab}\epsilon^{ij}\Gamma_{\mu}\epsilon_j(\text{Im}\,
{\cal N})_{IJ}Z^Je^{{\cal K}/2}\,, \label{deltapsi} \\
D_{\mu}(\omega)\epsilon^i &=& (\partial_{\mu}+\frac 14\omega^{ab}_{\mu}\Gamma_{ab})
\epsilon^i + \frac i2 A_{\mu}\epsilon^i + \partial_{\mu}q^X{\omega_{X\,j}}^i\epsilon^j
+ g A_{\mu}^I{P_{I\,j}}^i\epsilon^j\,, \label{derivata} \\
\delta\lambda^{\alpha}_i &=& -\frac 12 e^{{\cal K}/2}g^{\alpha\bar\beta}
{\cal D}_{\bar\beta}\bar Z^I(\text{Im}\,{\cal N})_{IJ}F^{-J}_{\mu\nu}\Gamma^{\mu\nu}
\epsilon_{ij}\epsilon^j + \Gamma^{\mu}{\cal D}_{\mu}z^{\alpha}\epsilon_i
+ gN^{\alpha}_{ij}\epsilon^j\,, \nonumber \\
\delta\zeta^A &=& \frac i2 f^{Ai}_X\Gamma^{\mu}{\cal D}_{\mu}q^X\epsilon_i +
g{\cal N}^{iA}\epsilon_{ij}\epsilon^j\,, \nonumber
\end{eqnarray}
where we defined
\begin{eqnarray}
S^{ij} &\equiv& -P^{ij}_Ie^{{\cal K}/2}Z^I\,, \nonumber \\
N^{\alpha}_{ij} &\equiv& e^{{\cal K}/2}\left[\epsilon_{ij}k^{\alpha}_I\bar Z^I -
2P_{Iij}{\cal D}_{\bar\beta}\bar Z^Ig^{\alpha\bar\beta}\right]\,, \qquad
{\cal N}^{iA} \equiv -if^{iA}_Xk^X_Ie^{{\cal K}/2}\bar Z^I\,. \nonumber
\end{eqnarray}
In \eqref{derivata}, $A_{\mu}$ is the gauge field of the K\"ahler U$(1)$,
\begin{equation}
A_{\mu} = -\frac i2(\partial_{\alpha}{\cal K}\partial_{\mu}z^{\alpha} -
\partial_{\bar\alpha}{\cal K}\partial_{\mu}\bar z^{\bar\alpha}) - gA^I_{\mu}P^0_I\,,
\label{KaehlerU1}
\end{equation}
with the moment map function
\begin{equation}
P^0_I = \langle T_I {\cal V}\,,\bar{\cal V}\rangle\,, \label{P0I}
\end{equation}
and
\begin{equation}
T_I{\cal V} \equiv\left(\begin{array}{cc} -{f_{IJ}}^K & 0 \\ C_{I,KJ} & {f_{IK}}^J
\end{array}\right)\left(\begin{array}{c} X^J \\ F_J\end{array}\right)\,.
\label{TIV}
\end{equation}
The major part of this paper will deal with the case of vector multiplets only,
i.~e.~, $n_H=0$. Then there are still two possible solutions of \eqref{equivariance}
for the moment maps $\vec P_I$, which are called SU(2) and U(1) Fayet-Iliopoulos
(FI) terms respectively \cite{Vambroes}. Here we are interested in the latter.
In this case
\begin{equation}
\vec P_I = \vec e\,\xi_I\,, \label{FI}
\end{equation}
where $\vec e$ is an arbitrary vector in SU(2) space and $\xi_I$ are constants
for the $I$ corresponding to U(1) factors in the gauge group.
If, moreover, we assume ${f_{IJ}}^K=0$ (abelian gauge group), and $k^{\alpha}_I=0$
(no gauging of special K\"ahler isometries), then only the $V_3$ part survives
in the scalar potential \eqref{scal-pot}, and one can also choose $C_{I,JK}=0$.
Note that this case corresponds to a gauging of a U(1) subgroup of the SU(2)
R-symmetry, with gauge field $\xi_IA^I_{\mu}$.

\section{Orbits of spinors under the gauge group}
\label{orbits}

A Killing spinor\footnote{Our conventions for spinors and their description in terms
of forms can be found in appendix \ref{spin-forms}.} can be viewed as an SU(2) doublet
$(\epsilon^1, \epsilon^2)$, where an upper index means that a spinor has positive
chirality. $\epsilon^i$ is related to the negative chirality spinor $\epsilon_i$ by
charge conjugation, $\epsilon_i^C = \epsilon^i$, with
\begin{equation}
\epsilon_i^C = \Gamma_0 C^{-1}\epsilon_i^{\ast}\,.
\end{equation}
Here $C$ is the charge conjugation matrix defined in appendix \ref{spin-forms}.
As $\epsilon^1$ has positive chirality, we can write $\epsilon^1 = c1 + de_{12}$
for some complex functions $c,d$. Notice that $c1 + de_{12}$ is in the same orbit as 1
under Spin(3,1), which can be seen from
\begin{displaymath}
e^{\gamma\Gamma_{13}}e^{\psi\Gamma_{12}}e^{\delta\Gamma_{13}}e^{h\Gamma_{02}}\,1
= e^{i(\delta+\gamma)}e^h\cos\psi\,1 + e^{i(\delta-\gamma)}e^h\sin\psi\,e_{12}\,.
\end{displaymath}
This means that we can set $c=1$, $d=0$ without loss of generality. In order to
determine the stability subgroup of $\epsilon^1$, one has to solve the
infinitesimal equation
\begin{equation}
\alpha^{cd}\Gamma_{cd}1 = 0\,, \label{stab}
\end{equation}
which implies $\alpha^{02} = \alpha^{13} = 0$, $\alpha^{01} = -\alpha^{12}$,
$\alpha^{03} = \alpha^{23}$. The stability subgroup of 1 is thus generated by
\begin{equation}
X = \Gamma_{01} - \Gamma_{12}\,, \qquad Y = \Gamma_{03} + \Gamma_{23}\,. \label{XY}
\end{equation}
One easily verifies that $X^2 = Y^2 = XY = 0$, and thus $\exp(\mu X
+ \nu Y) = 1 + \mu X + \nu Y$, so that $X,Y$ generate $\bR^2$.

Having fixed $\epsilon^1 = 1$, also $\epsilon_1$ is determined by
$\epsilon_1 = \epsilon^{1C} = e_1$. A negative chirality spinor independent of
$\epsilon^1$ is $\epsilon_2$, which can be written as a linear combination of
odd forms, $\epsilon_2 = ae_1 + be_2$, where $a$ and $b$ are again complex valued
functions. We can now act with the stability subgroup of $\epsilon^1$ to bring
$\epsilon_2$ to a special form:
\begin{displaymath}
(1 + \mu X + \nu Y)(ae_1 + be_2) = be_2 + [a - 2b(\mu + i\nu)]e_1\,.
\end{displaymath}
In the case $b=0$ this spinor is invariant, so the representative is
$\epsilon^1=1$, $\epsilon_2=ae_1$ (so that $\epsilon^2 = \bar a 1$), with isotropy
group $\bR^2$. If $b \neq 0$, one can bring the spinor to the form $be_2$
(which implies $\epsilon^2 = -\bar b e_{12}$), with isotropy group $\bI$.
The representatives\footnote{Note the difference in form compared to the Killing spinors
of the corresponding theories in five and six dimensions: in six dimensions these can be
chosen constant \cite{Gutowski:2003rg} while in five dimensions they are constant up to
an overall function \cite{Grover:2006ps}. In four dimensions such a choice is generically
not possible.} together with the stability subgroups are
summarized in table \ref{tab:orbits}.
Given a Killing spinor $\epsilon^i$, one can construct the bilinear
\begin{equation}
V_A = A(\epsilon^i,\Gamma_A\epsilon_i)\,,
\end{equation}
with the Majorana inner product $A$ defined in \eqref{Majorana}, and the sum
over $i$ is understood. For $\epsilon_2 = ae_1$, $V_A$ is lightlike, whereas for
$\epsilon_2 = be_2$ it is timelike, see table~\ref{tab:orbits}. The existence of a
globally defined Killing spinor $\epsilon^i$, with isotropy group $G \in$ Spin(3,1),
gives rise to a $G$-structure. This means that we have an
$\bR^2$-structure in the null case and an identity structure in the
timelike case.

In U(1) gauged supergravity, the local Spin(3,1) invariance is
actually enhanced to Spin(3,1) $\times$ U(1). For U(1) Fayet-Iliopoulos terms,
the moment maps satisfy \eqref{FI}, where we can choose $e^x = \delta^x_3$
without loss of generality. Then, under a gauge transformation
\begin{equation}
A^I_{\mu} \to A^I_{\mu} + \partial_{\mu}\alpha^I\,, \label{U(1)transf}
\end{equation}
the Killing spinor $\epsilon^i$ transforms as
\begin{equation}
\epsilon^1 \to e^{-ig\xi_I\alpha^I}\epsilon^1\,, \qquad
\epsilon^2 \to e^{ig\xi_I\alpha^I}\epsilon^2\,, \label{transf_eps}
\end{equation}
which can be easily seen from the supercovariant derivative
(cf.~eq.~\eqref{derivata}). Note that
$\epsilon^1$ and $\epsilon^2$ have opposite charges under the U(1).
In order to obtain the stability subgroup, one determines the Lorentz
transformations that leave the spinors $\epsilon^1$ and $\epsilon^2$ invariant
up to arbitrary phase factors $e^{i\psi}$ and $e^{-i\psi}$ respectively, which can
then be gauged away using the additional U(1) symmetry.
If $\epsilon_2=0$, one gets in this way an
isotropy group generated by $X, Y$ and $\Gamma_{13}$ obeying
\begin{displaymath}
[\Gamma_{13}, X] = -2Y\,, \qquad [\Gamma_{13}, Y] = 2X\,, \qquad [X, Y] = 0\,,
\end{displaymath}
i.~e.~$G\cong$ U(1)$\ltimes \bR^2$. For $\epsilon_2 = ae_1$ with $a \neq 0$,
the stability subgroup $\bR^2$ is not enhanced, whereas the $\bI$ of
the representative $(\epsilon^1,\epsilon_2) = (1,be_2)$ is promoted to U(1)
generated by $\Gamma_{13} = i\Gamma_{\bar\bullet\bullet}$. The Lorentz
transformation matrix $a_{AB}$ corresponding to $\Lambda =
\exp(i\psi\Gamma_{\bar\bullet\bullet}) \in$ U(1), with
$\Lambda\Gamma_B\Lambda^{-1} = {a^A}_B\Gamma_A$, has nonvanishing
components
\begin{equation}
a_{+-} = a_{-+} = 1\,, \qquad a_{\bullet\bar\bullet} = e^{2i\psi}\,, \qquad
a_{\bar\bullet\bullet} = e^{-2i\psi}\,. \label{residualU1}
\end{equation}
Finally, notice that in U(1) gauged supergravity one can choose the
function $a$ in $\epsilon_2=ae_1$ real and positive: Write $a =
R\exp(2i\delta)$, use
\begin{displaymath}
e^{\delta\Gamma_{13}}1 = e^{i\delta}1\,, \qquad
e^{\delta\Gamma_{13}}ae_1 = e^{-i\delta}ae_1 = e^{i\delta}Re_1\,,
\end{displaymath}
and gauge away the phase factor $\exp(i\delta)$ using the
electromagnetic U(1).

\begin{table}[ht]
\begin{center}
\begin{tabular}{||c||c|c||c||}
\hline $(\epsilon^1,\epsilon_2)$ & $G\subset$ Spin(3,1) & $G\subset$ Spin(3,1)
$\times$ U(1) &
$V_A E^A = A(\epsilon^i,\Gamma_A\epsilon_i)E^A$ \\
\hline\hline
$(1,0)$ & $\bR^2$ & U(1)$\ltimes \bR^2$$\,\,\,\,$ & $-\sqrt 2 E^-$ \\
\hline
$(1,ae_1)$ & $\bR^2$ & $\bR^2 \,\,\,\, (a \in \bR)$ & $-\sqrt 2(1+a^2)E^-$ \\
\hline
$(1,be_2)$ & $\bI$ & U(1)$\,\,\,\,$ &  $\sqrt 2(|b|^2E^+-E^-)$ \\
\hline
\end{tabular}
\end{center}
\caption{The representatives $(\epsilon^1,\epsilon_2)$ of the orbits of Weyl
spinors and their stability subgroups $G$ under the gauge groups
Spin(3,1) and Spin(3,1) $\times$ U(1) in the ungauged and U(1)-gauged
theories, respectively. The number of orbits is the same in both
theories, the only difference lies in the stability subgroups and
the fact that $a$ is real in the gauged theory. In the last column
we give the vectors constructed from the spinors.}\label{tab:orbits}
\end{table}

Note that in the gauged theory the presence of $G$-invariant Killing
spinors will in general not lead to a $G$-structure on the manifold
but to stronger conditions. The structure group is in fact reduced
to the intersection of $G$ with Spin(3,1), and hence is equal to
the stability subgroup in the ungauged theory.

The representatives, stability subgroups and vectors constructed from
the Killing spinors are summarized in table \ref{tab:orbits} both for
the ungauged and the U(1)-gauged cases.

\section{Null representative $(\epsilon^1,\epsilon_2)=(1,ae_1)$}
\label{null}

In this section we will analyze the conditions coming from a single
null Killing spinor, and determine all supersymmetric
solutions in this class. We shall first keep things general, i.~e.~,
including hypermultiplets and a general gauging, and write down the
linear system following from the Killing spinor equations. This system will
then be solved for the case of U(1) Fayet-Iliopoulos terms and without hypers,
while the solution in the general case will be left for a future
publication. As was explained before, it is always possible
to choose $a$ real and positive, so we shall set $a=e^{\chi}$ in what follows.

\subsection{Conditions from the Killing spinor equations}

From the vanishing of the hyperini variation one obtains
\begin{eqnarray}
\label{H1}\left(f_{X}^{1A}+e^\chi f_{X}^{2A}\right)\mathcal{D}_{+}q^X&=&0\ ,\\
\label{H2}\left(f_{X}^{1A}+e^\chi
f_{X}^{2A}\right)\mathcal{D}_{\bullet}q^X&=&ig\sqrt2\left(e^\chi
\mathcal{N}^{1A}-\mathcal{N}^{2A}\right)\ ,
\end{eqnarray}
whereas the gaugino variation yields
\begin{eqnarray}
-e^\chi e^{\mathcal{K}/2}g^{\alpha\overline{\beta}}\mathcal{D}_{\overline{\beta}}
\overline{Z}^I({\mathrm{Im}}\,\mathcal{N})_{IJ}(F^{-J+-}-F^{-J\bullet\overline
{\bullet}})&&\nonumber \\
+\sqrt2\mathcal{D}_{\bullet}z^{\alpha}+g(N^\alpha_{11}+e^\chi
N^\alpha_{12})&=&0\ , \label{G1}\\
\sqrt2 e^\chi e^{\mathcal{K}/2}g^{\alpha\overline{\beta}}\mathcal{D}_{\overline
{\beta}}\overline{Z}^I({\mathrm{Im}}\,\mathcal{N})_{IJ}F^{-J-\bullet}-
\mathcal{D}_+z^\alpha&=&0\ ,\label{G2} \\
e^{\mathcal{K}/2}g^{\alpha\overline{\beta}}\mathcal{D}_{\overline{\beta}}
\overline{Z}^I({\mathrm{Im}}\,\mathcal{N})_{IJ}(F^{-J+-}-F^{-J\bullet\overline
{\bullet}})&&\nonumber \\
+\sqrt2 e^\chi\mathcal{D}_{\bullet}z^{\alpha}+g(N^\alpha_{21}+e^\chi
N^\alpha_{22})&=&0\ ,\label{G3} \\
\sqrt2e^{\mathcal{K}/2}g^{\alpha\overline{\beta}}\mathcal{D}_{\overline{\beta}}
\overline{Z}^I({\mathrm{Im}}\,\mathcal{N})_{IJ}F^{-J-\bullet}+e^\chi
\mathcal{D}_+z^\alpha&=&0\ . \label{G4}
\end{eqnarray}
It is straightforward to show that the equations
(\ref{G1})-(\ref{G4}) imply that
\eq
\label{cga3}\mathcal{D}_+z^\alpha = 0\ ,
\feq
\eq
\label{cga1}\mathcal{D}_{\bullet}z^\alpha = -g\frac{N^\alpha_{11}+e^\chi
N^\alpha_{12}+e^\chi N^\alpha_{21}+e^{2\chi} N^\alpha_{22}}{\sqrt2
(1+e^{2\chi})}\ ,
\feq
\eq
\label{cga4}g^{\alpha\overline{\beta}}\mathcal{D}_{\overline{\beta}}\overline{Z}^I
({\mathrm{Im}}\,\mathcal{N})_{IJ}F^{-J-\bullet} = 0\ ,
\feq
\eq
\label{cga2}e^{\mathcal{K}/2}g^{\alpha\overline{\beta}}\mathcal{D}_{\overline
{\beta}}\overline{Z}^I({\mathrm{Im}}\,\mathcal{N})_{IJ}(F^{-J+-}-F^{-J\bullet
\overline{\bullet}}) = g\frac{e^{\chi}N^\alpha_{11}+e^{2\chi} N^\alpha_{12}-
N^\alpha_{21}-e^{\chi} N^\alpha_{22}}{1+e^{2\chi}}\ .
\feq
Finally, from the gravitini we get
\begin{eqnarray}
\label{Gr1}\omega^{+-}-\omega^{\bullet\overline{\bullet}}&=&2\sqrt2e^{\chi}
e^{\mathcal{K}/2}({\mathrm{Im}}\,\mathcal{N})_{IJ}Z^{J}F^{-I+\overline{\bullet}}
E^-\\
&&+2\sqrt 2 e^\chi\left[ge^{-\chi}S^{11}+gS^{12}-\frac{e^{\mathcal{K}/2}}2
({\mathrm{Im}}\,\mathcal{N})_{IJ}Z^{J}\left(F^{-I+-}-F^{-I\bullet\overline
{\bullet}}\right)\right]E^{\overline{\bullet}}\nonumber\\
&&-2\left(\mathcal{A}_1^{\ \ 1}+e^\chi\mathcal{A}_2^{\ \ 1}\right)-i A\ ,
\nonumber\\
\label{Gr2}\omega^{+-}-\omega^{\bullet\overline{\bullet}}&=&-2\sqrt2e^{-\chi}
e^{\mathcal{K}/2}({\mathrm{Im}}\,\mathcal{N})_{IJ}Z^{J}F^{-I+\overline{\bullet}}
E^-\\
&&+2\sqrt 2 e^{-\chi}\left[gS^{12}+ge^{\chi}S^{22}+\frac{e^{\mathcal{K}/2}}2
({\mathrm{Im}}\,\mathcal{N})_{IJ}Z^{J}\left(F^{-I+-}-F^{-I\bullet\overline
{\bullet}}\right)\right]E^{\overline{\bullet}}\nonumber\\
&&-2\left(\mathcal{A}_2^{\ \ 2}+e^{-\chi}\mathcal{A}_1^{\ \
2}\right)-i A\nonumber-2d\chi\ ,\nonumber\\
\label{Gr3}\omega^{-\bullet}&=&-\sqrt2e^\chi e^{\mathcal{K}/2}({\mathrm{Im}}\,
\mathcal{N})_{IJ}Z^{J}F^{-I-\bullet}E^{\overline{\bullet}}\\
&&+\sqrt2\left[gS^{11}+ge^\chi S^{12}+\frac{e^\chi e^{\mathcal{K}/2}}2
({\mathrm{Im}}\,\mathcal{N})_{IJ}Z^{J}\left(F^{-I+-}-F^{-I\bullet\overline
{\bullet}}\right)\right]E^-\ ,\nonumber\\
\label{Gr4}\omega^{-\bullet}&=&\sqrt2e^{-\chi}e^{\mathcal{K}/2}({\mathrm{Im}}\,
\mathcal{N})_{IJ}Z^{J}F^{-I-\bullet}E^{\overline{\bullet}}\\
&&+\sqrt2\left[ge^{-\chi}S^{12}+gS^{22}-\frac{e^{-\chi}e^{\mathcal{K}/2}}2
({\mathrm{Im}}\,\mathcal{N})_{IJ}Z^{J}\left(F^{-I+-}-F^{-I\bullet\overline
{\bullet}}\right)\right]E^-\ ,\nonumber
\end{eqnarray}
with the gauged SU(2) connection
\[
\mathcal{A}_i^{\ \ j}=gA^IP_{Ii}^{\ \ \
j}+dq^X\omega_{Xi}^{\ \ \ j}\ .
\]
From equations (\ref{Gr3}) and (\ref{Gr4}) one obtains
\begin{eqnarray}
\label{cgr1}e^{\mathcal{K}/2}({\mathrm{Im}}\,\mathcal{N})_{IJ}Z^{J}F^{-I-\bullet}&=&0
\ ,\\
\label{cgr2}e^{\mathcal{K}/2}
({\mathrm{Im}}\,\mathcal{N})_{IJ}Z^{J}\left(F^{-I+-}-F^{-I\bullet\overline{\bullet}}
\right)&=&-2g\frac{S^{11}+S^{12}(e^\chi-e^{-\chi})-S^{22}}{e^\chi+e^{-\chi}}\ ,
\end{eqnarray}
and
\begin{equation}\label{e1}
\omega^{-\bullet}=-\frac{\mathcal{C}_1}{\sqrt2}E^-\ ,
\end{equation}
with
\[
\mathcal{C}_1=-g\frac{e^{-\chi}S^{11}+2S^{12}+e^{\chi}S^{22}}{\cosh\chi}\ .
\]
As the $(n_V+1)\times(n_V+1)$ matrix $(Z^I,{\cal D}_{\bar\alpha}
{\bar Z}^I)$ is invertible \cite{Vambroes}, eqns.~\eqref{cgr1}, \eqref{cgr2}
together with \eqref{cga4}, \eqref{cga2} determine uniquely the fluxes
$F^{-I-\bullet}$ and $F^{-I+-}-F^{-I\bullet\bar\bullet}$, with the result\footnote{To
get this, one has to use \eqref{invImN}.}
\begin{eqnarray}
F^{-I-\bullet}&=&0\,, \nonumber \\
F^{-I+-}-F^{-I\bullet\overline{\bullet}} &=& 4g\frac{S^{11}+S^{12}(e^\chi
-e^{-\chi})-S^{22}}{e^\chi+e^{-\chi}}e^{\mathcal{K}/2}\overline{Z}^I \nonumber \\
&&-2g\frac{N^\alpha_{11}+e^{\chi} N^\alpha_{12}-e^{-\chi}N^\alpha_{21}-
N^\alpha_{22}}{e^{\chi}+e^{-\chi}}e^{\mathcal{K}/2}\mathcal{D}_\alpha Z^I\ .
\label{fluxes}
\end{eqnarray}
Moreover, antiselfduality
implies that
\begin{displaymath}
F^{-I+\bullet} = F^{-I-\bar\bullet} = F^{-I+-}+F^{-I\bullet\bar\bullet} = 0\,,
\end{displaymath}
so that all fluxes except $F^{-I+\bar\bullet}=:\psi^I$ are fixed.
Using \eqref{fluxes}, eqns.~(\ref{Gr1}) and (\ref{Gr2}) become
\begin{eqnarray}
\omega^{+-}-\omega^{\bullet\overline{\bullet}}&=&2\sqrt2e^{\chi}e^{\mathcal{K}/2}
({\mathrm{Im}}\,\mathcal{N})_{IJ}Z^J\psi^I E^-\nonumber \\
&&+2\sqrt2ge^\chi\left[\frac{(2+e^{-2\chi})S^{11}+2e^\chi
S^{12}-S^{22}}{e^\chi+e^{-\chi}}\right]E^{\overline{\bullet}}\nonumber\\
&&-2\left(\mathcal{A}_1^{\ \ 1}+e^\chi\mathcal{A}_2^{\ \ 1}\right)-i A\ ,\label{eqa}
\end{eqnarray}
\begin{eqnarray}
\omega^{+-}-\omega^{\bullet\overline{\bullet}}&=&-2\sqrt2e^{-\chi}e^{\mathcal{K}/2}
({\mathrm{Im}}\,\mathcal{N})_{IJ}Z^{J}\psi^I E^-\nonumber \\
&&+2\sqrt2ge^{-\chi}\left[\frac{2e^{-\chi}S^{12}-S^{11}+(2+e^{2\chi})S^{22}}{e^\chi+
e^{-\chi}}\right]E^{\overline{\bullet}}\nonumber\\
&&-2\left(\mathcal{A}_2^{\ \ 2}+e^{-\chi}\mathcal{A}_1^{\ \
2}\right)-iA-2d\chi\ ,\label{eqb}
\end{eqnarray}
from which one can determine some components of the spin connection
and the gauge potential $\cal A$ as follows:
First of all, \eqref{invImN} permits to decompose $\psi^I$ in a graviphoton
part $\psi$ and matter vector part $\psi^{\alpha}$ as
\eq
\psi^I = {\cal D}_{\alpha}X^I\psi^{\alpha} + i\bar X^I\psi\,, \label{decomp}
\feq
where
\eq
\psi^{\alpha} := -2g^{\alpha\bar\beta}{\cal D}_{\bar\beta}\bar X^J({\mathrm{Im}}\,
{\cal N})_{JK}\psi^K\,, \qquad i\psi:= -2X^J({\mathrm{Im}}\,{\cal N})_{JK}\psi^K\,.
\feq
Then, the sum of the real parts of (\ref{eqa}) and (\ref{eqb}) yields
\begin{eqnarray}
\label{E2}\omega^{+-}&=&\sqrt2\sinh\chi {\mathrm{Im}}\psi E^--
\frac{\overline{\mathcal{C}}_2}{\sqrt2}E^\bullet-\frac{\mathcal{C}_2}{\sqrt2}
E^{\overline{\bullet}}+2\sinh\chi\mbox{Re}\mathcal{A}_1^{\ \ 2}-d\chi\ ,
\end{eqnarray}
with
\eq
\mathcal{C}_2 = -g\frac{e^{\chi}S^{11}+(e^{2\chi}+e^{-2\chi})S^{12}+e^{-\chi}S^{22}}
{\cosh\chi}\ .\nonumber
\feq
On the other hand, the difference of the real parts of (\ref{eqa}) and (\ref{eqb})
gives
\eq
\label{E5}{\mathrm{Re}}\,\mathcal{A}_1^{\ \ 2}=\frac1{2\cosh\chi}\left(\frac{
\overline{\mathcal{C}}_3}{\sqrt2}E^\bullet+\frac{\mathcal{C}_3}{\sqrt2}
E^{\overline{\bullet}}-d\chi\right)-\frac1{\sqrt2}{\mathrm{Im}}\psi E^-\ ,
\feq
where we defined
\[
\mathcal{C}_3=-2g(S^{11}+2S^{12}\sinh\chi-S^{22})\ .
\]
Plugging (\ref{E5}) into (\ref{E2}) one gets
\eq
\omega^{+-}=-\frac{\overline{\mathcal{C}}_1}{\sqrt2}E^\bullet-\frac{\mathcal{C}_1}
{\sqrt2}E^{\overline{\bullet}}-\frac{e^\chi}{\cosh\chi}d\chi\ .
\feq
From the sum of the imaginary parts of (\ref{eqa}) and (\ref{eqb}) we
have
\eq
\label{E3}\omega^{\bullet\overline{\bullet}}=i\sqrt 2\sinh\chi{\mathrm{Re}}\psi
E^--\frac{\overline{\mathcal{C}}_2}{\sqrt2}E^\bullet
+\frac{\mathcal{C}_2}{\sqrt2}E^{\overline{\bullet}}+iA+2i\cosh\chi{\mathrm{Im}}
\mathcal{A}_1^{\ \ 2}\ .
\feq
Finally, the difference of the imaginary parts of (\ref{eqa}) and
(\ref{eqb}) yields
\begin{eqnarray}
\label{E4}\mathcal{A}_1^{\ \ 1}+i\sinh\chi\mbox{Im}\mathcal{A}_1^{\ \ 2}&=&
\frac1{2\sqrt 2}(\overline{\mathcal{C}}_3E^\bullet - \mathcal{C}_3
E^{\bar\bullet})-\frac i{\sqrt2}\cosh\chi{\mathrm{Re}}\psi E^-\ .
\end{eqnarray}
Summarizing, the components $\omega^{-\bullet}$, $\omega^{+-}$ and
$\omega^{\bullet\bar\bullet}$ are fixed by the supersymmetry conditions, while
the remaining components will be determined below by imposing vanishing torsion.

In order to obtain the spacetime geometry, we consider the spinor bilinears
\eq
V_{\mu\ j}^{\ i} = A(\epsilon^i, \Gamma_{\mu}\epsilon_j)\ ,
\feq
where the Majorana inner product is defined in \eqref{Majorana}. The
nonvanishing components are
\eq
V_{-\ 1}^{\ 1} = -\sqrt 2\ , \qquad V_{-\ 2}^{\ 2} = -e^{2\chi}\sqrt 2\ , \qquad
V_{-\ 2}^{\ 1} = V_{-\ 1}^{\ 2} = -e^{\chi}\sqrt 2\ .
\feq
This yields for the trace part
\eq
V_A E^A \equiv V_{A\ i}^{\ i}E^A = -\sqrt2(1+e^{2\chi})E^-\ .
\feq
Using the identities
\eq
\omega_{Xi}^{\ \ \ j\,\ast} = -\omega_{Xj}^{\ \ \ i}\ , \qquad P_{Ii}^{\ \ j\,\ast}
= -P_{Ij}^{\ \ i}\ ,
\feq
it is straightforward to shew that the linear system \eqref{Gr1} - \eqref{Gr4}
implies the following constraints:
\eq
\partial_+\chi + \frac12\omega_+^{+-}(1+e^{-2\chi}) = 0\ , \qquad
\partial_-\chi + \frac12\omega_-^{+-}(1+e^{-2\chi}) = 0\ , \qquad
\omega_+^{-\bullet} = 0\ , \nonumber
\feq
\eq
\partial_{\bullet}\chi + \frac12(\omega_{\bullet}^{+-}-\omega_-^{-\bar\bullet})
(1+e^{-2\chi}) = 0\ , \qquad \omega^{-\bar\bullet}_{\bullet} = 0\ , \qquad
\omega_{\bullet}^{-\bullet} + \omega_{\bar\bullet}^{-\bar\bullet} = 0\ .
\feq
These equations are easily shown to be equivalent to
\eq
\partial_A V_B + \partial_B V_A -\omega^C_{\ \ B|A}V_C -\omega^C_{\ \ A|B}V_C=0\ ,
\feq
(where $\omega^C_{\ \ B|A}=\omega^{CD}_A\eta_{DB}$), which means that $V$ is
Killing. Note that $V^2=0$, so $V$ is lightlike.

The next step is to impose zero torsion.
The torsion two-form reads
\begin{eqnarray}
T^+&=&dE^++E^+\wedge\left(\frac{\overline{\mathcal{C}_1}}{\sqrt2}E^{\bullet}+
\frac{\mathcal{C}_1}{\sqrt2}E^{\overline{\bullet}}
+\frac{e^\chi}{\cosh\chi}d\chi\right)
+\omega^{+\overline{\bullet}}\wedge E^{\bullet}
+\omega^{+\bullet}\wedge E^{\overline{\bullet}}\ ,\nonumber\\
T^-&=&dE^--E^-\wedge\left(\sqrt2\overline{\mathcal{C}}_1E^\bullet + \sqrt2
\mathcal{C}_1E^{\overline{\bullet}}+\frac{e^\chi}{\cosh\chi}d\chi\right)\ ,
\nonumber\\
T^{\bullet}&=&dE^{\bullet}+E^-\wedge\left(\frac{\mathcal{C}_1}{\sqrt2}E^+
+i\sqrt 2\sinh\chi{\mathrm{Re}}\psi E^{\bullet}+\omega^{+\bullet}\right)
\nonumber \\
&&\qquad -E^{\bullet}\wedge\left(\frac{\mathcal{C}_2}{\sqrt2}E^{\overline{\bullet}}
+ iA + 2i\cosh\chi{\mathrm{Im}}{\cal A}_1^{\ \ 2}\right)\ .\nonumber
\end{eqnarray}
From the vanishing of $T^-$ one gets $E^-\wedge dE^-=0$, so by
Fr\"{o}benius' theorem there exist two functions $H$ and
$u$ such that locally
\eq
E^-=\frac{du}H\ .\label{E^-}
\feq
Let us introduce a coordinate $v$ such that
\[
V=\frac{\partial}{\partial v}\ .
\]
Since $V$ is proportional to $E_+$ as a vector, and $\langle E_+,E^-\rangle=0$,
$u$ is independent of $v$, and thus can be used as a further coordinate.
Taking into account that
\begin{eqnarray}
\langle V,E^+\rangle&=&-\sqrt2(1+e^{2\chi})\langle E_+,E^+\rangle
=-\sqrt2(1+e^{2\chi})\ ,\nonumber\\
\langle V,E^{\bullet}\rangle&=&-\sqrt2(1+e^{2\chi})\langle E_+,E^\bullet\rangle
=0\ ,\nonumber
\end{eqnarray}
we obtain
\[
E^+_{\ \ v}=-\sqrt2(1+e^{2\chi})\ ,\qquad E^\bullet_{\ \
v}=E^{\overline{\bullet}}_{\ \ v}=0\ .
\]

Up to now, our discussion is completely general, i.~e.~, it includes
hypermultiplets and a general gauging.
In the remainder of this paper, we shall specialize to the case without
hypers and no gauging
of special K\"ahler isometries ($k^{\alpha}_I=0$). The inclusion of
hypermultiplets will be studied in a forthcoming publication.
This leaves two possible solutions for the moment maps \cite{Vambroes},
namely SU(2) or U(1) Fayet-Iliopoulos (FI) terms. We shall consider here the
latter, which satisfy \eqref{FI}, where $e^x=\delta^x_3$ without loss of
generality\footnote{$e^x=\delta^x_3$ can always be achieved by a global SU(2)
rotation (which is a symmetry of the theory).}. One has then
\begin{eqnarray}
&&P_{I1}^{\ \ 1}=-P_{I2}^{\ \ 2}=i\xi_I\ ,\ \ \ \ P_{I1}^{\ \ 2}=P_{I2}^{\ \ 1}=0\ ,
\nonumber\\
&&S^{12}=S^{21}=i\xi_IZ^Ie^{\mathcal{K}/2}\ ,\ \ \ S^{11}=S^{22}=0\ ,
\end{eqnarray}
\eq
N_{11}^\alpha=N_{22}^\alpha=0\ ,\ \ \ N_{12}^\alpha=N_{21}^{\alpha}=-2i\xi_I
e^{\mathcal{K}/2}\mathcal{D}_{\bar\beta}{\bar Z}^Ig^{\alpha\bar\beta}
\ ,\nonumber
\feq
as well as
\eq
\mathcal{A}_1^{\ \ 2}=\mathcal{A}_2^{\ \ 1}=0\ ,\qquad
\mathcal{A}_1^{\ \ 1}=-\mathcal{A}_2^{\ \ 2} = igA^I\xi_I
\feq
and $\mathcal{D}_\mu z^\alpha=\partial_\mu z^\alpha$. Equ.~\eqref{fluxes}
implies then for the fluxes
\eq
F^I=-2ig\tanh\chi(\mbox{Im}\,\mathcal{N})^{-1|IJ}\xi_J E^{\bullet}\wedge
E^{\bar\bullet}+\psi^I E^-\wedge E^{\bullet}+{\bar\psi}^I E^-\wedge E^{\bar\bullet}\ ,
\label{fluxesFI}
\feq
while \eqref{cga1} leads to the flow equation
\eq
\partial_\bullet z^\alpha=\frac{ig\sqrt2 e^{\mathcal{K}/2}}{\cosh\chi}
g^{\alpha\bar\beta}\mathcal{D}_{\bar\beta}{\bar Z}^I\xi_I \label{flow}
\feq
for the scalars.

Notice that the special U(1)$\ltimes\bR^2$ orbit with representative
$(\epsilon^1,\epsilon_2)=(1,0)$, that can be obtained in the limit
$\chi\to-\infty$, cannot occur in the FI case with nontrivial scalar fields:
Multiplying \eqref{eqb} with $e^{\chi}$ and letting $\chi\to-\infty$ yields
(with ${{\cal A}_1}^2=0$) $gS^{12}=0$, so that either $g=0$ (ungauged case) or
$\xi_IZ^I=0$, which implies $\xi_I{\cal D}_{\alpha}Z^I=0$, so that the scalars
are constant (cf.~section \ref{const-scal}). In the presence of hypermultiplets
and general gauging however, this orbit might occur, so there would be one more
representative to consider.

In order to proceed, it is convenient to distinguish two subcases, namely
$d\chi=0$ and $d\chi\neq 0$.

\subsection{Constant Killing spinor, $d\chi=0$}
\label{sec-chi=0}

If $d\chi=0$, equation (\ref{E5}) reduces to
\[
\frac{1}{2\cosh\chi}\left(\frac{\overline{\mathcal{C}}_3}{\sqrt2}E^\bullet
+\frac{\mathcal{C}_3}{\sqrt2}E^{\overline{\bullet}}\right)-\frac1{\sqrt 2}
{\mathrm{Im}}\psi E^-=0\ ,
\]
and thus ${\mathrm{Im}}\psi=0$ and $\mathcal{C}_3=0$, which implies $\chi=0$.
Let us denote the remaining two coordinates by $w,\bar w$ (with $\bar w$ the
complex conjugate of $w$) and define
${\cal G}\equiv {E^+}_u$, so that the null tetrad reads
\begin{eqnarray}
E^+&=&{\cal G}du-2\sqrt2 dv+E^+_{\ \ w}dw+E^+_{\ \ \bar w}d\bar w\ ,\nonumber\\
E^-&=&\frac{du}H\ ,\nonumber\\
E^{\bullet}&=& E^{\bullet}_{\ \ u}du+E^{\bullet}_{\ \ w}dw+E^{\bullet}_{\ \ \bar w}
d\bar w\ .\nonumber
\end{eqnarray}
To simplify $E^{\bullet}$, first perform a diffeomorphism
\[
w\mapsto w'(u,w,\bar w)
\]
obeying
\eq
{E^{\bullet}}_w\frac{\partial w'}{\partial\bar w} + {E^{\bullet}}_{\bar w}
\frac{\partial\bar w'}{\partial\bar w} = 0\ . \label{equ-w'}
\feq
This eliminates ${E^{\bullet}}_{\bar w}$. Notice that due to the
Cauchy-Kovalevskaya theorem, it is always possible to solve \eqref{equ-w'}
locally for $w'$\footnote{Because $\partial_v$ is Killing, ${E^{\bullet}}_w$
and ${E^{\bullet}}_{\bar w}$ can depend on $v$ only by a common phase factor
$e^{i\lambda(u,v,w,\bar w)}$, so that a potential $v$-dependence drops out of
\eqref{equ-w'}.}. Finally, the component ${E^{\bullet}}_u$ can be removed
using the residual gauge freedom, given by the stability subgroup
$\bR^2$ of the null spinor. To see this,
consider an $\bR^2$ transformation with group element
\[
\Lambda = 1 + \mu X + \nu Y\ ,
\]
where $X$ and $Y$ are given in \eqref{XY}. Defining $\alpha=\mu+i\nu$, this
can also be written as
\eq
\Lambda = 1 + \alpha\Gamma_{+\bullet} + \bar\alpha\Gamma_{+\bar\bullet}\ .
\label{alpha-transf}
\feq
Given the ordering $A,B=+,-,\bullet,\bar\bullet$, the Lorentz transformation
matrix $a_{AB}$ corresponding to $\Lambda\in\bR^2\subseteq{\mathrm{Spin}}(3,1)$
reads
\eq
a_{AB} = \left(\begin{array}{cccc} 0 & 1 & 0 & 0 \\
                                  1 & -4|\alpha|^2 & 2\bar\alpha & 2\alpha \\
                                  0 & -2\bar\alpha & 0 & 1 \\
                                  0 & -2\alpha & 1 & 0 \end{array}\right)\ .
\feq
The transformed vierbein $^{\alpha}E^A = {a^A}_BE^B$ is thus given by
\begin{eqnarray}
^{\alpha}E^+ &=& E^+ + 2\bar\alpha E^{\bullet} + 2\alpha E^{\bar\bullet}
- 4|\alpha|^2 E^-\ , \qquad ^{\alpha}E^- = E^-\ , \nonumber \\
^{\alpha}E^{\bullet} &=& E^{\bullet} - 2\alpha E^-\ , \qquad
^{\alpha}E^{\bar\bullet} = E^{\bar\bullet} - 2\bar\alpha E^-\ . \label{transf-E}
\end{eqnarray}
Choosing $\alpha={E^{\bullet}}_u/2{E^-}_u$ eliminates ${E^{\bullet}}_u$, so that
we can take
\[
E^{\bullet} = {E^{\bullet}}_wdw\ , \qquad E^{\bar\bullet} =
{E^{\bar\bullet}}_{\bar w}d\bar w
\]
without loss of generality. Then the inverse tetrad reads
\eq
E_+ = -\frac1{2\sqrt2}\partial_v\ , \qquad E_- = H(\partial_u + \frac{{\cal G}}
{2\sqrt2}\partial_v)\ , \qquad E_{\bullet} = \frac1{{E^{\bullet}}_w}(\partial_w
+ \frac{{E^+}_w}{2\sqrt2}\partial_v)\ .
\feq
In what follows we shall set ${E^{\bullet}}_w\equiv \rho e^{i\zeta}$.

Equ.~\eqref{E4} reduces to
\eq
gA^I\xi_I = -\frac1{\sqrt2}\psi E^-\ , \label{xiAchi=0}
\feq
while \eqref{cga3} and \eqref{cga1} lead to
\eq
\partial_v z^{\alpha} = 0 \label{dvz=0}
\feq
and
\eq
\partial_w z^{\alpha} = ig\sqrt2\xi_Ie^{\mathcal{K}/2}\mathcal{D}_{\bar\beta}
{\bar Z}^Ig^{\alpha\bar\beta}\rho e^{i\zeta} \label{dwz}
\feq
respectively. \eqref{dvz=0} implies that the scalars are independent of $v$
and thus $A_v=0$.
The vanishing of the torsion gives the missing components $\omega^{+\bullet}$
of the spin connection (that we do not list here), plus the additional
constraints
\begin{eqnarray}
\label{FI1}2i(A_u + \partial_u\zeta) &=& -\frac1{H\rho^2}({E^+}_{w,\bar w}-
{E^+}_{\bar w,w})\ ,\\
\label{FI2}\partial_w\ln H &=& 2\sqrt2 ig\,\xi_I{\bar Z}^Ie^{\mathcal{K}/2}\rho
e^{i\zeta}\ ,\\
\label{FI3}\partial_v\rho &=& \partial_v\zeta = 0\ ,\\
\label{FI4}i(A_{\bar w} + \partial_{\bar w}\zeta) &=& \sqrt2 ig\,\xi_IZ^I
e^{\mathcal{K}/2}\rho e^{-i\zeta} - \partial_{\bar w}\ln\rho
\end{eqnarray}
on the null tetrad.

All that remains to be done at this point is to impose the Bianchi identities
and the Maxwell equations, which read respectively
\[
dF^I=0\ , \qquad d\mbox{Re}(\mathcal{N}_{IJ}F^{+J})=0\ .
\]
The fluxes can be obtained by setting $\chi=0$ in \eqref{fluxesFI},
\begin{eqnarray}
F^I &=& \psi^IE^-\wedge E^\bullet+{\bar\psi}^IE^-\wedge E^{\bar\bullet}\nonumber \\
    &=& \frac{du}H\wedge(\psi^I\rho e^{i\zeta}dw + {\bar\psi}^I\rho e^{-i\zeta}
        d\bar w)\ ,
\end{eqnarray}
with the selfdual part
\eq
F^{+I} = {\bar\psi}^IE^-\wedge E^{\bar\bullet}\ .
\feq
One finds that the Bianchi identities imply $\partial_v\psi^I=0$ and
\eq
\partial_{\bar w}(\psi^I\rho e^{i\zeta}/H) =
\partial_w({\bar\psi}^I\rho e^{-i\zeta}/H)\ , \label{Bianchichi=0}
\feq
whereas the Maxwell equations give $\partial_v{\cal N}_{IJ}=0$ (which is
automatically satisfied since $\partial_v z^{\alpha}=0$) and
\eq
\partial_{\bar w}({\bar{\cal N}}_{IJ}\psi^J\rho e^{i\zeta}/H) =
\partial_w({\cal N}_{IJ}{\bar\psi}^J\rho e^{-i\zeta}/H)\ . \label{Maxwchi=0}
\feq
Note that imposing $dF^I=0$ is actually not sufficient; we must also ensure
that $\xi_IF^I=\xi_IdA^I$, because the linear combination $\xi_IA^I$ is
determined by the Killing spinor equations (cf.~\eqref{xiAchi=0}). This
leads to the additional condition
\eq
g\sqrt2\xi_I\psi^I\rho e^{i\zeta} = H\partial_w\left(\frac{\psi}H\right)\ .
\label{F=dAchi=0}
\feq
In conclusion, the null tetrad (with the exception of ${E^+}_u={\cal G}$),
gauge fields and scalars are determined by the
coupled system \eqref{dvz=0}-\eqref{FI4} and
\eqref{Bianchichi=0}-\eqref{F=dAchi=0}.
Finally, the wave profile $\cal G$ is fixed by the $uu$ component
of the Einstein equations, which are given in (C.1) of \cite{Cacciatori:2008ek},
where in our case
\begin{eqnarray}
R_{uu} &=& -\frac{2{\cal G}_{,w\bar w}}{H\rho^2} - \frac{({E^+}_{w,\bar w} -
          {E^+}_{\bar w,w})^2}{2H^2\rho^4} + \frac{{E^+}_{w,u\bar w} - {E^+}_{\bar w,
          uw}}{H\rho^2}\nonumber \\
      & & -2\partial^2_u\ln\rho - 2(\partial_u\ln\rho)(\partial_u\ln(H\rho))
          \nonumber \\
      & & +\frac1{H\rho^2}\left[({\cal G}_{,\bar w} - {E^+}_{\bar w,u})\partial_w
          \ln H + ({\cal G}_{,w} - {E^+}_{w,u})\partial_{\bar w}\ln H\right]
          \nonumber \\
      & & + \frac{2{\cal G}}{H\rho^2}(\partial_w\partial_{\bar w}\ln H - (\partial_w
          \ln H)(\partial_{\bar w}\ln H))\ .
\end{eqnarray}
Then, as was shown in \cite{Cacciatori:2008ek}, all other equations of motion
are automatically satisfied.

Notice that the above system is K\"ahler-covariant, as it must be:
Under a K\"ahler transformation
\eq
{\cal K} \to {\cal K} + f(z^{\alpha}) + {\bar f}({\bar z}^{\bar\alpha})\ ,
\label{Kaehlertransf}
\feq
the Killing spinors transform as
\eq
\epsilon^i \to e^{\frac14(\bar f - f)}\epsilon^i\ , \qquad
\epsilon_i \to e^{-\frac14(\bar f - f)}\epsilon_i\ .
\feq
In order for our representative $(\epsilon^1,\epsilon_2)=(1,e_1)$ to be
invariant, this must be compensated by a Spin$(3,1)$ transformation
$\Lambda=\exp((\bar f-f)\Gamma_{\bullet\bar\bullet}/4)$. The corresponding
matrix $a_{AB}\in$ SO$(3,1)$ is given in \eqref{residualU1}, from which
we see that $E^{\bullet}$ takes a phase factor $\exp(-i(\bar f-f)/2)$,
so that $\zeta$ is shifted according to
\eq
\zeta \to \zeta - \frac i2(\bar f - f)\ . \label{shiftzeta}
\feq
Taking into account \eqref{Kaehlertransf}, \eqref{shiftzeta}, as well as
\[
Z^I \to Z^Ie^{-f}\ , \qquad \psi^I = F^{-I+\bar\bullet}\to {a^+}_A
{a^{\bar\bullet}}_B F^{-IAB} = \psi^Ie^{-\frac12(\bar f - f)}\ ,
\]
it is easy to show that the system \eqref{dvz=0}-\eqref{FI4} and
\eqref{Bianchichi=0}-\eqref{F=dAchi=0} is K\"ahler-covariant.

In what follows, we shall obtain explicit solutions in some special cases.

\subsection{Explicit solutions for $d\chi=0$}

\subsubsection{Constant scalars}
\label{const-scal}

If we assume
\eq
\xi_I{\cal D}_{\alpha}Z^I=0\ , \label{DZ=0}
\feq
the flow equation \eqref{dwz} implies $\partial_wz^{\alpha}=0$. Actually,
since the scalar potential $V=g^2V_3$ satisfies
\eq
\partial_{\bar\beta}V = 4g^2\xi_I\xi_J\left[g^{\gamma\bar\delta}e^{{\cal K}}
{\cal D}_{\gamma}Z^I{\cal D}_{\bar\beta}{\cal D}_{\bar\delta}{\bar Z}^J -
2e^{{\cal K}}Z^J{\cal D}_{\bar\beta}{\bar Z}^I\right]\ , \label{abl_V}
\feq
\eqref{DZ=0} forces the scalars to be constant, i.~e., they do not depend on
$\bar w$ and $u$ either\footnote{This is true if the potential has no flat
directions.}.
One has then $A_{\mu}=0$, so that \eqref{FI4} and
the complex conjugate of \eqref{FI2} give
\eq
\partial_{\bar w}(\ln\rho + i\zeta + \frac12\ln H) = 0\ ,
\feq
and thus
\eq
\rho e^{i\zeta}\sqrt H = f(u,w)\ , \label{f(u,w)}
\feq
with $f(u,w)$ an arbitrary function. Defining $F(u,w)$ by
$f=\partial_w F$ we get
\eq
E^{\bullet} = H^{-1/2}\partial_w F dw = H^{-1/2}\left[dF - \frac{\partial F}
             {\partial u}du\right]\ .
\feq
By a diffeomorphism $w'=F(u,w)$ combined with a local Lorentz transformation
\eqref{alpha-transf} to eliminate ${E^{\bullet}}_u$ one can thus set (after
dropping the primes) $E^{\bullet}=H^{-1/2}dw$ without loss of generality, so
that $\rho=H^{-1/2}$, $\zeta=0$. From \eqref{FI1} one obtains
\[
{E^+}_{w,\bar w} - {E^+}_{\bar w,w} = 0\ ,
\]
and hence ${E^+}_w=\partial_w m$ for some real function $m$ that can always
be set to zero by shifting $v$ and $\cal G$. Finally, \eqref{FI2} yields
\eq
\sqrt H = {\cal C}w + \bar{\cal C}\bar w + A(u)\ ,
\feq
where we defined the constant
\[
{\cal C} = \sqrt2 ig\,\xi_I{\bar Z}^Ie^{{\cal K}/2}\ ,
\]
and $A(u)$ is an arbitrary real function. The metric has the form of a
Lobachevski wave on AdS,
\eq
ds^2 = \frac 2H[{\cal G}du^2 - 2\sqrt 2 dvdu + dwd\bar w]\ .
\feq
Note that, by shifting $w$, $\cal G$ and $v$ appropriately, one can always
achieve $A(u)=0$.

To obtain the gauge fields, observe that the Bianchi identities
\eqref{Bianchichi=0} imply that $\partial_w({\bar\psi}^IH^{-3/2})$ is real,
\[
\partial_w({\bar\psi}^IH^{-3/2}) = \lambda^I(u,w,\bar w)\ , \qquad \lambda^I =
{\bar\lambda}^I\ .
\]
From the Maxwell equations \eqref{Maxwchi=0} one concludes that
${\cal N}_{IJ}\lambda^J$ must be real as well, and thus
\eq
({\mathrm{Im}}\,{\cal N})_{IJ}\lambda^J = 0\ .
\feq
As ${\mathrm{Im}}\,{\cal N}$ is invertible, this yields $\lambda^I=0$,
so that
\eq
\psi^I = H^{3/2}\rho^I(u,w)\ ,
\feq
for some function $\rho^I(u,w)$. Taking into account \eqref{DZ=0},
equ.~\eqref{decomp} gives
\[
\psi = \frac{\sqrt 2 g}{\cal C}\xi_I\psi^I = \frac{\sqrt 2 g}{\cal C}H^{3/2}
       \xi_I\rho^I\ .
\]
Since $\psi$ must be real, this implies
that the linear combination $\xi_I\rho^I$ can depend on $u$ only.
Then, one easily shows that \eqref{F=dAchi=0} is automatically satisfied.
Analogous to \eqref{decomp}, we can decompose
\eq
\rho^I = {\cal D}_{\alpha}X^I\rho^{\alpha} + i{\bar X}^I\rho\ , \label{dec-rho}
\feq
where (using $\xi_I\rho^I = i\xi_I{\bar X}^I\rho$)
\eq
\rho = \rho(u)\ , \qquad \rho^{\alpha} = \rho^{\alpha}(u,w)\ . \label{rho}
\feq
The fluxes are thus given by
\eq
F^I = \rho^I du\wedge dw + {\bar\rho}^I du\wedge d\bar w\ ,
\feq
with $\rho^I$ satisfying \eqref{dec-rho} and \eqref{rho}. Note that this
solution with constant scalars includes the one in minimal gauged
supergravity found in \cite{Caldarelli:2003pb}.

\subsubsection{Prepotential $F=-iZ^0Z^1$}
\label{simple-prepot}

We now consider a simple model determined by the prepotential
\eq
F = -iZ^0Z^1\ ,
\feq
that has $n_V=1$ (one vector multiplet), and thus just one complex scalar $z$.
Choosing $Z^0=1$, $Z^1=z$ (cf.~\cite{Vambroes}), the symplectic vector $v$ reads
\eq
v = \left(\begin{array}{c} 1 \\ z \\ -iz \\ -i\end{array}\right)\ .
\feq
The K\"ahler potential, metric and kinetic matrix for the vectors are given
respectively by
\eq
e^{-{\cal K}} = 2(z + \bar z)\ , \qquad g_{z\bar z} = \partial_z\partial_{\bar z}
{\cal K} = (z + \bar z)^{-2}\ ,
\feq
\eq
{\cal N} = \left(\begin{array}{cc} -iz & 0 \\ 0 & -\frac iz\end{array}\right)\ .
\feq
Note that positivity of the kinetic terms in the action requires
${\mathrm{Re}}z>0$. For the scalar potential one obtains
\eq
V = g^2V_3 = -\frac{4g^2}{z+\bar z}(\xi_0^2 + 2\xi_0\xi_1z + 2\xi_0\xi_1\bar z
+ \xi_1^2z\bar z)\ , \label{pot-simple-model}
\feq
which has an extremum at $z=\bar z=|\xi_0/\xi_1|$. In what follows we assume
$\xi_I>0$. The K\"ahler U(1) is
\eq
A_{\mu} = \frac i{2(z+\bar z)}\partial_{\mu}(z-\bar z)\ .
\feq
In order to solve the system \eqref{dvz=0}-\eqref{FI4} and
\eqref{Bianchichi=0}-\eqref{F=dAchi=0} we shall take $z=\bar z$ (this includes
the extremum of the potential, and thus the AdS vacuum) and
$\zeta={E^+}_w={E^+}_{\bar w}=\psi^I=0$. Then, $A_{\mu}=0$, and the only
nontrivial equations are
\eqref{dwz}, \eqref{FI2} and \eqref{FI4}, which become
\begin{eqnarray}
\partial_wz &=& \sqrt2ig\sqrt z(-\xi_0 + \xi_1z)\rho\ , \label{dwz-modellino} \\
\partial_w\ln H &=& \sqrt2 ig\frac{\xi_0+\xi_1z}{\sqrt z}\rho\ , \label{dwH} \\
\partial_{\bar w}\ln\rho &=& ig\frac{\xi_0+\xi_1z}{\sqrt{2z}}\rho\ .
\label{dbarwrho}
\end{eqnarray}
\eqref{dwH} and the complex conjugate of \eqref{dbarwrho} can be combined to
give
\eq
\partial_w\ln(\rho\sqrt H) = 0\ ,
\feq
and hence
\eq
\rho\sqrt H = g(u,\bar w)\ ,
\feq
where $g(u,\bar w)$ denotes an arbitrary function. Because $\rho\sqrt H$ is real,
$g(u,\bar w)$ can depend only on $u$. As was explained in section
\ref{const-scal}, one can set $g(u)=1$ without loss of generality by a
combination of a diffeomorphism $w\to w/g(u)$ and a local Lorentz
transformation \eqref{alpha-transf}, so that $\rho=H^{-1/2}$.
From \eqref{dwz-modellino} and \eqref{dwH} we get
\eq
H = \frac{(-\xi_0+\xi_1z)^2}zf(u)\ , \label{H(z)}
\feq
with $f(u)$ an arbitrary function that we will take equal to one in the following.
Since $z$ is real, \eqref{dwz-modellino} yields $\partial_xH=0$, where we
introduced the real coordinates $x,y$ by $w=x+iy$. Let us further assume that
also $\partial_uH=0$, so that $H$ (and thus, by virtue of \eqref{H(z)}, also
$z$) depends only on $y$. Then, the flow equation \eqref{dwz-modellino}
together with $\rho=H^{-1/2}$ and \eqref{H(z)} implies
\eq
z = \frac{\xi_0}{\xi_1}e^{-2\sqrt2gy}\ , \label{z(y)}
\feq
where the integration constant was chosen such that the scalar goes to
its critical value for $y\to0$. The metric becomes
\eq
ds^2 = \frac1{2\xi_0\xi_1\sinh^2\!\sqrt2gy}\left[{\cal G}du^2 - 2\sqrt2\,dudv + dx^2
       + dy^2\right]\ , \label{domainwall}
\feq
where $\cal G$ is determined by the $uu$ component of the Einstein equations.
\eqref{domainwall} describes a gravitational wave propagating
on a domain wall. For $z\to\xi_0/\xi_1$ ($y\to0$), the geometry becomes that of a
wave on AdS$_4$.

Note that perhaps some of the assumptions made
above (like reality of $z$) can be relaxed while maintaining integrability
of the equations. Another possible generalization is the inclusion of
nonvanishing gauge fields. This will be done in section \ref{1/2-dchi0}.

\subsubsection{Ungauged case}

Finally let us check if we correctly reproduce the results of
\cite{Meessen:2006tu} in the ungauged case. If $g=0$, the flow equation
\eqref{dwz} implies $\partial_w z^{\alpha}=0$, and thus
$z^{\alpha}=z^{\alpha}(\bar w,u)$. Using \eqref{KaehlerU1}, this gives
\[
A_{\bar w} = -\frac i2\partial_{\bar w}{\cal K}\ ,
\]
so that \eqref{FI4} leads to
\eq
\partial_{\bar w}(\ln\rho + i\zeta + \frac12{\cal K}) = 0\ .
\feq
This can be integrated to give
\eq
\rho e^{i\zeta} = e^{-\frac12{\cal K} + h(w,u)}\ ,
\feq
with $h(w,u)$ an arbitrary function that can be set to zero without loss
of generality by a combination of a diffeomorphism $w\to w'(u,w)$ and
a local Lorentz transformation \eqref{alpha-transf}.
From \eqref{FI2} we have $\partial_wH=\partial_{\bar w}H=0$, and hence $H=H(u)$
so that we can set $H=1$ by a redefinition of $u$.

In conclusion, the metric is given by
\eq
ds^2 = 2du\left[{\cal G}du - 2\sqrt2 dv + {E^+}_w dw + {E^+}_{\bar w}d\bar w\right]
       +2e^{-{\cal K}}dwd\bar w\ , \label{metricg=0}
\feq
where
\eq
2iA_u = -e^{{\cal K}}({E^+}_{w,\bar w} - {E^+}_{\bar w,w})\ . \label{E+g=0}
\feq
The scalars are arbitrary functions of $\bar w,u$, and the gauge fields read
\eq
F^I = e^{-{\cal K}/2}\psi^I du\wedge dw + e^{-{\cal K}/2}{\bar\psi}^I
du\wedge d\bar w\ , \label{Fg=0}
\feq
with $\psi^I$ determined by
\begin{eqnarray}
\partial_{\bar w}(e^{-{\cal K}/2}\psi^I) &=&
\partial_w(e^{-{\cal K}/2}{\bar\psi}^I)\ , \nonumber \\
\partial_{\bar w}(e^{-{\cal K}/2}{\bar{\cal N}}_{IJ}\psi^J) &=&
\partial_w(e^{-{\cal K}/2}{\cal N}_{IJ}{\bar\psi}^J)\ . \label{Maxg=0}
\end{eqnarray}
\eqref{metricg=0}, \eqref{E+g=0}, \eqref{Fg=0} and \eqref{Maxg=0} exactly
coincide with the equations obtained in \cite{Meessen:2006tu} that
admit pp-waves and cosmic strings as solutions. Because \eqref{xiAchi=0}
implies in addition $\psi=0$, one actually gets only a subclass of the solutions
of \cite{Meessen:2006tu}. As $\chi=0$ is not the only case to consider, this is
not surprising.

\subsection{Killing spinor with $d\chi\neq 0$}
\label{dchineq0}

In the case $d\chi\neq 0$ we can determine explicitely the function $H$ appearing
in \eqref{E^-}: From equ.~(\ref{E5}) one has
\eq \label{ne1}
d\chi = -\sqrt2\cosh\chi{\mathrm{Im}}\psi E^-+2ig\sqrt2\sinh\chi e^{\mathcal{K}/2}
\xi_I({\bar Z}^IE^{\bullet}-Z^IE^{\bar\bullet})\ .
\feq
Plugging this into $T^-=0$ we obtain
\[
d\left[\left(e^{2\chi}-1\right)E^-\right]=0\ ,
\]
and therefore one can introduce a function $u$ such that
\eq
\left(e^{2\chi}-1\right)E^-=du\ \ \Rightarrow\ \
E^-=\frac{du}{e^{2\chi}-1}\ . \label{E^-chi}
\feq
On the other hand, \eqref{E4} gives
\eq \label{ne2}
g\texttt{A}=-\frac1{\sqrt2}\cosh\chi{\mathrm{Re}}\psi E^-+g\sqrt2\sinh\chi
e^{\mathcal{K}/2}\xi_I({\bar Z}^IE^{\bullet}+Z^IE^{\bar\bullet})\ ,
\feq
where we defined
\[
\texttt{A}=A^I\xi_I\ .
\]
(\ref{ne1}) and (\ref{ne2}) determine the components $E^{\bullet}$,
$E^{\bar\bullet}$ of the null tetrad,
\begin{eqnarray}
E^{\bullet}&=&-\frac1{2\sqrt2g\sinh\chi{\bar S}^{12}}\left(\frac{i\bar\psi
\coth\chi}{2\sqrt2e^\chi}du+\frac{d\chi}2+ig\texttt{A}\right)\ ,\nonumber \\
E^{\bar\bullet}&=&-\frac1{2\sqrt2g\sinh\chi S^{12}}\left(-\frac{i\psi\coth\chi}
{2\sqrt2e^\chi}du+\frac{d\chi}2-ig\texttt{A}\right)\ .\nonumber
\end{eqnarray}
We already introduced the coordinates $u$, $v$. Using \eqref{ne1} together
with $V=\partial_v=-\sqrt2(1+e^{2\chi})E_+$, we get
$\langle\partial_v,d\chi\rangle=0$, and thus $\partial\chi/\partial v=0$, so
that $\chi$ is independent of $v$. Furthermore, \eqref{ne1} and \eqref{E^-chi}
imply that $du\wedge d\chi\neq 0$, therefore the function $\chi$ must depend
nontrivially on the two remaining coordinates. This allows to choose $\chi$ as
a further coordinate. Finally, the fourth coordinate will be called $\Psi$.
Notice that due to $\langle V,E^{\bullet}\rangle=0$, $\texttt{A}$ has no
$v$-component, $\texttt{A}_v=0$.

Now we employ the $\mathbb{R}^2$ stability subgroup of the null
spinor (cf.~\eqref{transf-E}) to set ${E^\bullet}_u={E^{\bar\bullet}}_u=0$.
This amounts to the choice
\eq
\frac{\psi\coth\chi}{2\sqrt2 e^{\chi}} + g\texttt{A}_u = 0\ , \label{gaugecond}
\feq
and hence ${\mathrm{Im}}\psi=0$. Using also
\eq
E^+ = {\cal G}du - \sqrt2(1+e^{2\chi})dv + {E^+}_{\chi}d\chi + {E^+}_{\Psi}d\Psi\ ,
\feq
we can proceed to impose vanishing torsion. $T^{\bullet}=0$ determines the
following components of the spin connection:
\begin{eqnarray}
\omega^{\bullet +}_v &=& 4g e^{\chi}S^{12}\ , \nonumber \\
\omega^{\bullet +}_{\chi} &=& -\frac{\sqrt2 gS^{12}}{\cosh\chi}{E^+}_{\chi}
+ 2e^{\chi}\sinh\chi(\partial_u + iA_u + \frac i{\sqrt2}e^{-\chi}
\psi){E^{\bullet}}_{\chi}\ , \nonumber \\
\omega^{\bullet +}_{\Psi} &=& -\frac{\sqrt2 gS^{12}}{\cosh\chi}{E^+}_{\Psi}
+ 2e^{\chi}\sinh\chi(\partial_u + iA_u + \frac i{\sqrt2}e^{-\chi}
\psi){E^{\bullet}}_{\Psi}\ ,
\end{eqnarray}
whereas $T^+=0$ gives
\[
\omega^{\bullet +}_u = \frac{8ig\sinh^2\!\chi {|S^{12}|}^2}{\texttt{A}_{\Psi}}
\left[{E^{\bullet}}_{\Psi}({E^+}_{u,\chi} - {E^+}_{\chi,u}) - {E^{\bullet}}_{\chi}
({E^+}_{u,\Psi} - {E^+}_{\Psi,u})\right] - \frac{\sqrt2g S^{12}e^{2\chi}}{\cosh\chi}
{E^+}_u\ ,
\]
together with the constraint
\begin{eqnarray}
\lefteqn{{E^+}_{\chi,\Psi} - e^{\chi}\cosh\chi\partial_{\chi}\left(\frac{{E^+}_{\Psi}}
{e^{\chi}\cosh\chi}\right)} \nonumber \\
&& + \frac{\texttt{A}_{\Psi}\psi}
{2\sqrt2 g\sinh\chi S^{12}{\bar S}^{12}} - 2e^{\chi}\sinh\chi\epsilon^{mn}
({E^{\bullet}}_m{\cal D}_uE_{\bullet n} + {E^{\bar\bullet}}_m{\cal D}_u
E_{\bar\bullet n}) = 0\ , \label{E^+chiPsi}
\end{eqnarray}
that determines ${E^+}_{\chi}$ and ${E^+}_{\Psi}$. In \eqref{E^+chiPsi} we
introduced the indices $m,n,\ldots=\chi,\Psi$,
and the convention $\epsilon^{\chi\Psi}=1$. The
K\"ahler-covariant derivatives ${\cal D}_u$ appearing in \eqref{E^+chiPsi} are
defined as
\[
{\cal D}_u {E^{\bullet}}_n = (\partial_u+iA_u){E^{\bullet}}_n\ , \qquad
{\cal D}_u {E^{\bar\bullet}}_n = (\partial_u-iA_u){E^{\bar\bullet}}_n\ .
\]
(As we remarked in section \ref{sec-chi=0}, in order for the spinor representative
to be invariant under a K\"ahler transformation, one must compensate with
a Spin(3,1) transformation, which acts also on $E^{\bullet},E^{\bar\bullet}$).

Finally we have to ensure that the Maxwell equations and Bianchi identities
hold. From $\xi_IF^I=\xi_IdA^I$ we obtain
\begin{eqnarray}
\partial_v\texttt{A}_u&=&0\ , \qquad \partial_{\chi}\texttt{A}_{\Psi} -
\partial_{\Psi}\texttt{A}_{\chi} = -(\mbox{Im}\,\mathcal{N})^{-1|IJ}\xi_I\xi_J
\frac{\texttt{A}_{\Psi}}{2{|S^{12}|}^2\sinh2\chi}\ , \label{FchiPsi} \\
\label{FdA4}\xi_I\psi^I&=&\frac{2\sqrt2e^\chi\sinh^2\chi}{\texttt{A}_\Psi}
\left\{\texttt{A}_{\Psi,u}-\texttt{A}_{u,\Psi}\right.\\
&&\hspace{3.1cm}\left.+2ig\left[\texttt{A}_\chi\left(\texttt{A}_{u,\Psi}-
\texttt{A}_{\Psi,u}\right)-\texttt{A}_\Psi\left(\texttt{A}_{u,\chi}-
\texttt{A}_{\chi,u}\right)\right]\right\}\Xb\ .\nonumber
\end{eqnarray}
Imposing the Bianchi identities, $dF^I=0$, one gets $\partial_v\psi^I=0$ and
\begin{eqnarray}
\frac1{\sinh2\chi}\partial_u\left[\frac{(\mbox{Im}\,\mathcal{N})^{-1|IJ}\xi_J}
{\X\,\Xb}\texttt{A}_\Psi\right]
+\frac1{\sqrt2}\partial_\chi\left[\frac{\texttt{A}_\Psi}{e^\chi\sinh^2\!\chi}
\mbox{Re}\left(\frac{\psi^I}{\Xb}\right)\right]&&\nonumber\\
-\frac1{\sqrt2e^\chi\sinh^2\!\chi}\partial_\Psi\left[\texttt{A}_\chi\mbox{Re}
\left(\frac{\psi^I}{\Xb}\right)+\frac1{2g}\mbox{Im}
\left(\frac{\psi^I}{\Xb}\right)\right]&=&0\ . \label{B}
\end{eqnarray}
The Maxwell equations $d\mbox{Re}\left(\mathcal{N}_{IJ}F^{+J}\right)=0$ yield
in addition
\begin{eqnarray}
\xi_I\bigg[{E^+}_{\chi,\Psi}&-& e^{\chi}\cosh\chi\partial_{\chi}
\left(\frac{{E^+}_{\Psi}}{e^{\chi}\cosh\chi}\right)\bigg] = \nonumber \\
&-&\frac{e^\chi}{4g\sinh\chi}\partial_u\left[\frac{\left(\mbox{Re}\,\mathcal{N}
\right)_{IJ}\left(\mbox{Im}\,\mathcal{N}\right)^{-1|JL}\xi_L}
{\X\,\Xb}\texttt{A}_\Psi\right]\nonumber\\
&-&\frac{e^\chi\cosh\chi}{2\sqrt2g}\partial_\chi\left[\frac{\texttt{A}_\Psi}{e^\chi
\sinh^2\!\chi}\mbox{Re}\left(\frac{\bar{\mathcal{N}}_{IJ}\psi^J}{\Xb}
\right)\right]\nonumber\\
&+&\frac{\cosh\chi}{2\sqrt2g\sinh^2\!\chi}\partial_\Psi\left[\texttt{A}_\chi
\mbox{Re}\left(\frac{\bar{\mathcal{N}}_{IJ}\psi^J}{\Xb}\right)
+\frac1{2g}\mbox{Im}\left(\frac{\bar{\mathcal{N}}_{IJ}\psi^J}{\Xb}
\right)\right].\label{M}
\end{eqnarray}
Notice that \eqref{cga3} implies $\partial_vz^{\alpha}=0$.
Using this together with the fact that $\partial_v$ is Killing, one
easily shows that all components of the vierbein do not depend on $v$ either.

The flow equation \eqref{flow} becomes
\eq
{\bar S}^{12}\sinh2\chi\left[\left(g\texttt{A}_{\chi}+\frac i2\right)
\partial_{\Psi} - g\texttt{A}_{\Psi}\partial_{\chi}\right]z^{\alpha} = ig
\texttt{A}_{\Psi}e^{{\cal K}/2}g^{\alpha\bar\beta}{\cal D}_{\bar\beta}{\bar Z}^I
\xi_I\ . \label{flow1}
\feq
In conclusion, the coupled system \eqref{E^+chiPsi}, \eqref{FchiPsi},
\eqref{FdA4}, \eqref{B}, \eqref{M} and \eqref{flow1} determines the
components of the null tetrad (except ${E^+}_u={\cal G}$), the functions
$\psi^I$ and the scalar fields $z^{\alpha}$.
The fluxes $F^I$ are then given by \eqref{fluxesFI}. Finally, the wave profile
$\cal G$ is fixed by the $uu$ component of the Einstein equations
(cf.~(C.1) of \cite{Cacciatori:2008ek}), where in our case
\begin{eqnarray}
R_{uu}&=&\frac{{E^+}_u{E^\bullet}_{\Psi}{E^{\bar\bullet}}_{\Psi}}{\sigma^2e^\chi
\sinh^3\!\chi\cosh^2\!\chi}+\frac{{E^+}_u\mbox{Re}\left[\left({E^\bullet}_{\chi,
\Psi}-{E^\bullet}_{\Psi,\chi}\right){E^{\bar\bullet}}_{\Psi}\right]}{\sigma^2e^\chi
\sinh^2\!\chi\cosh\chi} \nonumber \\
&&+\frac2{\sigma^2}|{E^{\bullet}}_{\Psi,u}{E^\bullet}_{\chi}-{E^{\bullet}}_{\chi,u}
{E^\bullet}_{\Psi}|^2 - \frac{\Upsilon^2}{8\sigma^2e^{2\chi}\sinh^2\!\chi}
\nonumber \\
&&+\frac{{E^{\bar\bullet}}_{\Psi}\Phi}{\sigma^2e^\chi\sinh\chi\sinh2\chi}
+\frac{\bar\Phi}{\sigma^2e^\chi\sinh\chi}\left({E^\bullet}_{\chi,\Psi}-{E^\bullet}_
{\Psi,\chi}+\frac{1+2e^{2\chi}}{\sinh2\chi}{E^\bullet}_{\Psi}\right) \nonumber \\
&&-\frac1{e^\chi\sinh\chi}\left[\frac12\partial_u\left(\frac{\Upsilon}{\sigma}
\right)+\frac{{E^\bullet}_{\Psi}}{\sigma e^{\chi}\sinh\chi}\partial_\chi
\left(\frac{e^\chi\sinh\chi\bar\Phi}{\sigma}\right)-\frac{{E^\bullet}_{\chi}}
{\sigma}\partial_\Psi\left(\frac{\bar\Phi}{\sigma}\right)\right] \nonumber \\
&&+\frac{2{E^+}_u}{e^\chi\sinh\chi}\mbox{Re}\left[\frac{{E^\bullet}_{\chi}}{\sigma
\sinh2\chi}\partial_\Psi\left(\frac{{E^{\bar\bullet}}_{\Psi}}{\sigma}\right)-
\frac{{E^\bullet}_{\Psi}}{\sigma}\partial_\chi\left(\frac{{E^{\bar\bullet}}_{\Psi}}
{\sigma\sinh2\chi}\right)\right]\ , \label{Ruu}
\end{eqnarray}
and we defined
\begin{eqnarray}
\sigma &=& {E^{\bullet}}_{\chi}{E^{\bar\bullet}}_{\Psi}-{E^{\bar\bullet}}_{\chi}
{E^\bullet}_{\Psi}\ ,\nonumber\\
\Phi&=&\left({E^+}_{u,\Psi}-{E^+}_{\Psi,u}\right){E^\bullet}_{\chi} + \left[
{E^+}_{\chi,u} - e^{\chi}\cosh\chi\partial_{\chi}\left(\frac{{E^+}_u}{e^{\chi}
\cosh\chi}\right)\right]{E^{\bullet}}_{\Psi}\ ,\nonumber \\
\Upsilon&=&{E^+}_{\chi,\Psi}-e^{\chi}\cosh\chi\partial_{\chi}\left(\frac{{E^+}_{\Psi}}
{e^{\chi}\cosh\chi}\right) + 2e^\chi\sinh\chi\partial_u\sigma\ .\nonumber
\end{eqnarray}
Then, as was shown in \cite{Cacciatori:2008ek}, all other equations of motion
are automatically satisfied. Note that $R_{uu}$ in \eqref{Ruu} can be rewritten
in a manifestly real form, but then the expression becomes considerably longer.

In the next subsection we shall obtain an explicit solution to the above
equations.

\subsection{Explicit solutions for $d\chi\neq 0$}
\label{expl-dchineq0}

If one sets $\psi^I=\texttt{A}_u=\texttt{A}_{\chi}={E^+}_{\chi}={E^+}_{\Psi}=0$
and $z^{\alpha}=z^{\alpha}(\chi)$, $\texttt{A}_{\Psi}=\texttt{A}_{\Psi}(\chi)$,
the only nontrivial equations are \eqref{FchiPsi} and \eqref{flow1}, which
reduce to
\begin{eqnarray}
\sinh2\chi\partial_{\chi}\ln\texttt{A}_{\Psi} &=& -\frac{(\mbox{Im}\,
\mathcal{N})^{-1|IJ}\xi_I\xi_J}{2{|S^{12}|}^2}\ , \label{A_Psi} \\
\sinh2\chi\partial_{\chi}z^{\alpha} &=& -\frac{ie^{{\cal K}/2}g^{\alpha\bar\beta}
{\cal D}_{\bar\beta}{\bar Z}^I\xi_I}{{\bar S}^{12}}\ . \label{flow2}
\end{eqnarray}
Note that \eqref{flow2} can also be written in the form
\eq
\sinh2\chi\partial_{\chi}z^{\alpha} = g^{\alpha\bar\beta}\partial_{\bar\beta}W\ ,
\feq
with the superpotential $W=\ln(\xi_I{\bar Z}^I)+{\cal K}$.

In what follows, we solve \eqref{A_Psi} and \eqref{flow2} for the simple
model with prepotential $F=-iZ^0Z^1$ introduced in section \ref{simple-prepot}.
Assuming in addition that the single scalar $z=Z^1$ is real, \eqref{A_Psi}
and \eqref{flow2} become
\begin{displaymath}
\sinh2\chi\partial_\chi z=-2z\frac{\xi_0-\xi_1z}{\xi_0+\xi_1z}\ , \qquad
\sinh2\chi\partial_\chi\ln\texttt{A}_\Psi=2\frac{\xi_0^2+\xi_1^2z^2}
{\left(\xi_0+\xi_1z\right)^2}\ ,
\end{displaymath}
with the solution
\eq
z_{\pm} = \frac{\xi_0}{\xi_1}+c\tanh\chi\pm\sqrt{c\tanh\chi\left(\frac{2\xi_0}
{\xi_1}+c\tanh\chi\right)}\ , \label{zpm}
\feq
\eq
\texttt{A}_{\Psi}^{\pm} = \tilde c\left(\frac{\xi_0^2}{z_{\pm}} - \xi_1^2z_{\pm}
\right)\ ,
\feq
where $c,\tilde c$ are integration constants. Finally, for the metric and the
nonvanishing components of the fluxes one obtains respectively
\begin{eqnarray}
ds^2&=&\frac{{\cal G}du^2}{e^\chi\sinh\chi}-2\sqrt2\coth\chi
dudv \nonumber \\
&& +\frac{z_\pm}{\sinh^2\!\chi(\xi_0+\xi_1z_\pm)^2}\left[\frac{d\chi^2}{4g^2}+
(\texttt{A}_\Psi^\pm)^2d\Psi^2\right]\ , \label{bubble}
\end{eqnarray}
\eq
F^0_{\chi\Psi} = \frac{2\xi_0\tilde c(\xi_0-\xi_1z)}{z(\xi_0+\xi_1z)\sinh2\chi}\ ,
\qquad F^1_{\chi\Psi} = \frac{2\xi_1\tilde cz(\xi_0-\xi_1z)}{(\xi_0+\xi_1z)
\sinh2\chi}\ .
\feq
Notice that $z_{\pm}$ in \eqref{zpm} are related by the strong-weak coupling
duality
\eq
z \to \frac{\xi_0^2}{\xi_1^2z}\ , \label{duality}
\feq
that sends $z_+$ to $z_-$ and vice versa. \eqref{duality} is actually a
residual $\bZ_4$ symmetry that remains of the full symplectic duality group
Sp$(4,\bR)$ after the gauging: In the notation of \cite{Vambroes}, it
corresponds to
\eq
{\cal S} = \left(\begin{array}{cc} A & B \\ C & D\end{array}\right)\in
{\mbox{Sp}}(4,\bR)\ ,
\feq
with $A=D=0$,
\eq
C = \left(\begin{array}{cc} \xi_0/\xi_1 & 0 \\ 0 & \xi_1/\xi_0
\end{array}\right)\ ,
\feq
and $B=-C^{-1}$. Since ${\cal S}^2=-\bI$, this generates $\bZ_4$.
Note also that the scalar potential \eqref{pot-simple-model} as well as the
vacuum values $z=\xi_0/\xi_1$ are invariant under \eqref{duality}.

Let us now briefly discuss the properties of the spacetime \eqref{bubble}.
Introducing the new radial coordinate $\rho=\sqrt{\coth\chi}$, we have
$\rho\to\infty$ for $\chi\to 0_+$ and $\rho\to1$ for $\chi\to\infty$.
Asymptotically for $\rho\to\infty$ one has $z_{\pm}\to\xi_0/\xi_1$ (the vacuum),
and the metric approaches
\eq
ds^2 \to \rho^2\left[{\cal G}du^2 - 2\sqrt2dudv + 2\xi_1^2{\tilde c}^2cd\Psi^2
\right] + \frac{d\rho^2}{8g^2\xi_0\xi_1\rho^2}\ ,
\feq
which represents a Lobachevski wave on AdS$_4$. On the other hand, for
$\rho\to1$, $z_+\to2c$, $z_-\to\xi_0^2/(2c\xi_1^2)$, $\texttt{A}_{\Psi}^{\pm}$
goes to a constant, and
\begin{displaymath}
\frac{z_{\pm}}{\sinh^2\!\chi(\xi_0+\xi_1z_{\pm})^2}\left[\frac{d\chi^2}{4g^2}
+\left(\texttt{A}_{\Psi}^{\pm}\right)^2d\Psi^2\right] \to \frac{z_{\pm}}
{2g^2(\xi_0+\xi_1z_{\pm})^2}\left[dR^2 + 4g^2R^2\left(\texttt{A}_{\Psi}^{\pm}
\right)^2d\Psi^2\right]\ ,
\end{displaymath}
where we defined $R={\mbox{arcosh}}\rho$. From this it is evident that in order for
the metric to be regular at $\rho=1$, one must identify\footnote{The
requirement that $g_{uu}$ behaves well at $\rho=1$ puts some additional
constraints on $\cal G$.}
\begin{displaymath}
\Psi \sim \Psi + \frac{\pi}{g|\texttt{A}_{\Psi}^{\pm}|_{\rho=1}}\ .
\end{displaymath}
As the spacetime ends at $\rho=1$, \eqref{bubble} can be interpreted as a
(wave on a) bubble of nothing \cite{Witten:1981gj,Aharony:2002cx} that
asymptotes to (a wave on) AdS$_4$. Notice that, in order to have a well-defined
limit for the case of constant scalars ($c=0$), one must choose
$\tilde c\propto c^{-1/2}$.

\section{Half-supersymmetric backgrounds}
\label{half-susy}

Let us now investigate the additional conditions satisfied by half-supersymmetric
vacua. Again, we will do this separately for $d\chi\neq 0$ and $d\chi=0$.
As the stability subgroup of the first Killing spinor was already used,
the second one cannot be simplified anymore, and is thus of the general form
\[
\epsilon^1=a1+be_{12}\ ,\qquad \epsilon^2=c1+de_{12}\ ,\qquad
\epsilon_1=\bar a e_1-\bar b e_2\ ,\qquad \epsilon_2=\bar c e_1-\bar d
e_2\ ,
\]
where $a,b,c,d$ are complex-valued functions.

\subsection{Case $d\chi\neq 0$}

From $\delta\psi_+^i=0$ one obtains
\begin{eqnarray}
\label{v1}\partial_v a&=&4ige^\chi\left(e^\chi\bar d X^I-b\bar
X^I\right)\xi_I\ ,\\
\label{v2}\partial_v b&=&0\ ,\\
\label{v3}\partial_v c&=&4ige^\chi\left(e^{-\chi}\bar b X^I-d\bar
X^I\right)\xi_I\ ,\\
\label{v4}\partial_v d&=&0\ ,
\end{eqnarray}
while $\delta\psi_-^i=0$ leads to (using also (\ref{v1})-(\ref{v4}))
\begin{eqnarray}
\label{u1}\partial_ua&=&\frac{i\left(a-e^{-\chi}\bar
c\right)\psi}{2\sqrt2\sinh\chi} -\frac{ig\sqrt2e^\chi\left(\bar d
X^I-e^\chi b \bar
X^I\right)\xi_IE^+_{\ \ u}}{\cosh\chi}\\
&&+2igb\sqrt2\sinh\chi\Xb\left[E^+_{\ \ \chi,u}-E^+_{\ \
u,\chi}+\frac{i}{\texttt{A}_\Psi}
\left(\frac{1}{2g}-i\texttt{A}_\chi\right) \left(E^+_{\ \
u,\Psi}-E^+_{\ \ \Psi,u}\right)\right]\ ,\nonumber\\
\label{u2}\partial_u b&=&-\frac{i e^{-\chi}}{2\sinh\chi}\left[
2be^\chi\sinh\chi A_u-\frac{b\psi e^{-\chi}}{\sqrt2}+\frac{g\sqrt2\left(e^{-\chi}
\bar c-a\right)\X}{\cosh\chi}\right]\ ,\\
\label{u3}\partial_uc&=&\frac{ie^{-\chi}\left(\bar a-e^{-\chi}
c\right)\psi}{2\sqrt2\sinh\chi} -\frac{ig\sqrt2\left(e^{-\chi}\bar b
X^I-e^{2\chi} d \bar
X^I\right)\xi_IE^+_{\ \ u}}{\cosh\chi}\\
&&+2igd\sqrt2\sinh\chi\Xb\left[E^+_{\ \ \chi,u}-E^+_{\ \
u,\chi}+\frac{i}{\texttt{A}_\Psi}
\left(\frac{1}{2g}-i\texttt{A}_\chi\right) \left(E^+_{\ \
u,\Psi}-E^+_{\ \ \Psi,u}\right)\right]\ ,\nonumber\\
\label{u4}\partial_u d&=&-\frac{i e^{-\chi}}{2\sinh\chi}\left[
2de^\chi\sinh\chi A_u+\frac{d\psi e^{\chi}}{\sqrt2}+\frac{g\sqrt2\left(e^{\chi}\bar
a-c\right)\X}{\cosh\chi}\right]\ .
\end{eqnarray}
The integrability conditions of the system (\ref{v1})-(\ref{u4}) imply that
\begin{eqnarray}
\label{c}
c&=&e^\chi \bar a-\frac{\tau e^\chi}{g\sqrt2}\frac{\X}{\Xb}
\bar b\ , \\
\label{d}
d&=&e^{-\chi}\frac{\X}{\Xb}\bar b\ ,
\end{eqnarray}
where we defined
\eq
\tau=\frac{\cosh\chi}{\X}\left[-\frac{\psi e^{-\chi}}{\sqrt2}-2A_u
+i\partial_u\ln\left(\frac{\X}{\Xb}\right)\right]\ . \label{tau}
\feq
Plugging (\ref{d}) into (\ref{v1}) and (\ref{v3}) one gets
$\partial_v a=\partial_v c=0$ so that $a$, $b$, $c$ and $d$ are
functions of $u$, $\chi$ and $\Psi$ only. Using
(\ref{c}) and (\ref{d}) into (\ref{u1})-(\ref{u4}) we
see that if $b\neq0$ we have to impose\footnote{One easily shows that for
$b=0$, the second Killing spinor coincides (up to a constant prefactor) with
the first one, and thus is not linearly independent. In what follows we shall
therefore assume $b\neq 0$.}
\begin{eqnarray}
\label{kse7}\partial_u\tau&=&i\coth\chi\left[-\frac{\psi e^{-\chi}}{\sqrt2}
-A_u+\frac{ie^{-\chi}}{2\cosh\chi}\partial_u\ln\left(\frac{\X}{\Xb}\right)\right]\tau\\
&&-4ig^2e^{-\chi}\sinh2\chi\Xb\left[\frac{e^\chi E^+_{\ \
u}}{\cosh\chi}+E^+_{\ \ \chi,u}-E^+_{\ \
u,\chi}\right.\nonumber\\
&&\hspace{1.6cm}\left.-\frac{i}{\texttt{A}_\Psi}
\left(\frac{\tanh\chi}{2g}+i\texttt{A}_\chi\right) \left(E^+_{\ \
u,\Psi}-E^+_{\ \ \Psi,u}\right)\right]\ .\nonumber
\end{eqnarray}
Making use of \eqref{c} and \eqref{d}, the remaining gravitino variations
$\delta\psi^i_{\bullet}=\delta\psi^i_{\bar\bullet}=0$ reduce to
\begin{eqnarray}
\partial_\chi a&=&\frac{ie^\chi}{\sqrt2\X}
\left(\frac{1}{2g}-i\texttt{A}_\chi\right) \left[i\coth\chi
A_u+\frac{e^\chi}{2\sinh\chi}\partial_u\ln
\left(\frac{\X}{\Xb}\right)\right]b\nonumber \\
&&+\frac{e^\chi}{\sqrt2}\partial_u\left(\frac{\texttt{A}_\chi}{\X}\right)b
-\frac{e^\chi}{2g\sqrt2\X}\left(\frac{\psi e^{-2\chi}}{\sqrt2\sinh\chi}+
i\partial_u\ln\X\right)b\ , \label{achi} \\
\partial_\Psi a&=&\frac{e^\chi \texttt{A}_\Psi}{\sqrt2
\X}\left[i\coth\chi A_u+\frac{e^\chi}{2\sinh\chi}\partial_u\ln
\left(\frac{\X}{\Xb}\right)\right]b\nonumber \\
&&+\frac{e^\chi}{\sqrt2}\partial_u\left(\frac{\texttt{A}_\Psi}{\X}
\right)b\ , \label{aPsi} \\
\label{bchi}\partial_\chi b&=&-\left(\coth2\chi+iA_\chi\right)b\ , \qquad
\partial_\Psi b=-iA_\Psi b\ ,
\end{eqnarray}
together with
\eq
2iA_m=\partial_m\ln\left(\frac{\Xb}{\X}\right)\ , \qquad
m=\chi,\Psi\ , \label{A_m}
\feq
and
\begin{eqnarray}
\label{tauchi}\partial_\chi
\tau&=&\left(2\coth2\chi+iA_\chi+2ig\texttt{A}_\chi\right)\tau
-\frac{2ig\texttt{A}_\chi}{\X}\left[-\frac{\psi e^{-\chi}}{\sqrt2}\cosh\chi
-\cosh\chi A_u\right.\\
&&\left.+\frac{i}{2}\left(e^{-\chi}\partial_u\ln\X
-e^\chi\partial_u\ln\Xb\right)
+i\sinh\chi\partial_u\ln \texttt{A}_\Psi\right]\nonumber\\
&&+\frac{1}{\X}\left[\frac{\psi e^{-\chi}}
{2\sqrt2}\left(2\sinh\chi-\frac1{\sinh\chi}\right) +\sinh\chi
A_u\right.\nonumber\\
&&\left.+\frac{i}{2}\left(e^{-\chi}\partial_u\ln\X
+e^\chi\partial_u\ln\Xb\right)
+2g\sinh\chi\left(\texttt{A}_{\chi,u}-\texttt{A}_\chi\partial_u\ln
\texttt{A}_\Psi\right)\right]\ ,\nonumber
\end{eqnarray}
\begin{eqnarray}
\label{tauPsi}\partial_\Psi\tau&=&i\left(A_\Psi+2g\texttt{A}_\Psi\right)\tau-
\frac{2ig\texttt{A}_\Psi}{\X}\left[-\frac{\psi e^{-\chi}}{\sqrt2}\cosh\chi
-\cosh\chi A_u\right.\\
&&\left.+\frac{i}{2}\left(e^{-\chi}\partial_u\ln\X-e^\chi\partial_u\ln\Xb\right)
+i\sinh\chi\partial_u\ln \texttt{A}_\Psi\right]\nonumber\ .
\end{eqnarray}
The vanishing of the gaugino variations yields the additional conditions
\begin{eqnarray}
&&\label{gam1}g^{\alpha\bar\beta}\mathcal{D}_{\bar\beta}\bar
X^I\left[\frac{i \xi_I \tau}{\cosh\chi} -e^{-\chi}\sqrt2
\left(\mbox{Im}\,\mathcal{N}\right)_{IJ}\psi^J\right]=0\ ,\\
&&\label{gam3}\partial_{\bar\bullet}z^\alpha=-\frac{ig\sqrt2e^{\mathcal{K}/2}
g^{\alpha\bar\beta}\mathcal{D}_{\bar\beta}\bar
Z^I\xi_I}{\cosh\chi}\frac{\X}{\Xb}\ ,
\end{eqnarray}
as well as $\partial_u z^{\alpha}=0$. This implies $A_u=0$, so that \eqref{tau}
simplifies to
\[
\tau=-\frac{\psi e^{-\chi}\cosh\chi}{\sqrt2\X}
= 2g\texttt{A}_ue^{\chi}\tanh\chi\ ,
\]
where in the last step we used the gauge condition \eqref{gaugecond}.
Thus, equ.~(\ref{gam1}) reduces to
\begin{equation}
\label{gamm1}g^{\alpha\bar\beta}\mathcal{D}_{\bar\beta}\bar
X^I\left[\frac{\psi\xi_I}{\X}
-2i\left(\mbox{Im}\,\mathcal{N}\right)_{IJ}\psi^J\right]=0\ .
\end{equation}
Because $\psi=2i\left(\mbox{Im}\mathcal{N}\right)_{IJ}X^I\psi^J$, we have
moreover
\begin{equation}
\label{gammm1}X^I\left[\frac{\psi\xi_I}{\X}
-2i\left(\mbox{Im}\,\mathcal{N}\right)_{IJ}\psi^J\right]=0\ .
\end{equation}
Since the $(n_V+1)\times(n_V+1)$ matrix $(X^I,\mathcal{D}_{\bar\alpha}{\bar X}^I)$
is invertible, (\ref{gamm1}) and (\ref{gammm1}) imply
\begin{equation}
\psi^I=-\frac{i\psi\left(\mbox{Im}\,\mathcal{N}\right)^{-1|IL}\xi_L}{2\X}\ .
\label{psi^I}
\end{equation}
\eqref{gam3}, together with \eqref{flow}, leads to
\eq
\partial_\chi z^\alpha = \frac{g^{\alpha\bar\beta}\mathcal{D}_{\bar\beta}
\bar Z^I\xi_I}{\sinh2\chi\bar Z^J\xi_J}\ , \qquad
\partial_{\Psi} z^{\alpha} = 0\ , \label{z-chi-Psi}
\feq
so that the scalars are functions of $\chi$ only, and hence $A_\Psi$
vanishes as well. Notice that the relations \eqref{A_m} are identically
satisfied if \eqref{z-chi-Psi} hold.
The complex equation \eqref{kse7} boils down to
\begin{eqnarray}
\label{Auu}\partial_u\texttt{A}_u&=&-2e^{-\chi}\sinh\chi\X\,\Xb
\frac{E^+_{\ \ u,\Psi}-E^+_{\ \ \Psi,u}}{\texttt{A}_\Psi}\ ,\\
\label{Au}\texttt{A}_u^2&=&2e^{-\chi}\cosh\chi\X\,\Xb
\left[\frac{e^\chi E^+_{\ \ u}}{\cosh\chi}+E^+_{\ \
\chi,u}-E^+_{\ \
u,\chi}\right.\\
&&\hspace{4.2cm}\left.+\frac{\texttt{A}_\chi}{\texttt{A}_\Psi}\left(E^+_{\
\ u,\Psi}-E^+_{\ \ \Psi,u}\right)\right]\nonumber\ ,
\end{eqnarray}
while (\ref{tauchi}) and (\ref{tauPsi}) yield
\begin{eqnarray}
\label{tauchis}\partial_\chi\texttt{A}_u - \partial_u\texttt{A}_\chi &=&
-(\mbox{Im}\,\mathcal{N})^{-1|IJ}\xi_I\xi_J\frac{\texttt{A}_u}{2\sinh2\chi
\X\,\Xb}\ ,\\
\label{tauPsis}\partial_\Psi \texttt{A}_u - \partial_u\texttt{A}_\Psi &=& 0\ .
\end{eqnarray}
Using \eqref{psi^I} and \eqref{tauPsis}, it is easy to show that \eqref{tauchis}
is equivalent to \eqref{FdA4} that follows from $\xi_IF^I=\xi_IdA^I$.
Moreover, the Bianchi identities \eqref{B} are automatically satisfied
once \eqref{psi^I} and \eqref{tauPsis} hold. Note also
the similarity between \eqref{tauchis} and \eqref{FchiPsi}.
Equ.~\eqref{E^+chiPsi} becomes
\begin{eqnarray}
\lefteqn{{E^+}_{\chi,\Psi} - e^{\chi}\cosh\chi\partial_{\chi}\left(\frac{{E^+}_{\Psi}}
{e^{\chi}\cosh\chi}\right)}\nonumber \\
&& - \frac{\texttt{A}_{\Psi}\texttt{A}_u e^{\chi}}{\cosh\chi\X\,\Xb}
- \frac{e^{\chi}}{2\sinh\chi\X\,\Xb}\left[\texttt{A}_{\chi}
\texttt{A}_{\Psi,u} - \texttt{A}_{\Psi}\texttt{A}_{\chi,u}\right] = 0\ .
\label{E^+final}
\end{eqnarray}
Making use of this, together with \eqref{FchiPsi}, \eqref{tauchis} and
\eqref{tauPsis}, one shows that the Maxwell equations \eqref{M} are identically
satisfied.

\eqref{FchiPsi}, \eqref{tauchis} and \eqref{tauPsis} can be easily integrated,
with the result
\eq
\texttt{A}_{\mu} = (\X\,\Xb\tanh\chi)^{1/2}\partial_{\mu}
\Xi(u,\chi,\Psi)\ ,
\feq
where $\Xi(u,\chi,\Psi)$ denotes an arbitrary function obeying
$\partial_{\Psi}\Xi\neq 0$\footnote{$\texttt{A}_{\Psi}=0$ would lead to a
singular metric.}. Furthermore, \eqref{Auu}, \eqref{Au} and \eqref{E^+final}
imply for $E^+$
\begin{eqnarray}
\frac{{E^+}_u}{e^{\chi}\cosh\chi} - \frac{\texttt{A}_u^2}{2\X\,\Xb
\sinh\!2\chi} &=& \partial_u\Lambda\ , \nonumber \\
\frac{{E^+}_m}{e^{\chi}\cosh\chi} - \frac{\texttt{A}_u\texttt{A}_m}
{\X\,\Xb\sinh\!2\chi} &=& \partial_m\Lambda\ , \qquad
m = \chi, \Psi\ ,
\end{eqnarray}
with $\Lambda(u,\chi,\Psi)$ again a function that can be chosen at will.
By shifting $v$ one can set $\Lambda=0$ without loss
of generality. Finally, we may employ the residual gauge freedom related to
the choice of the coordinate $\Psi$ that consists in sending
$\Psi\mapsto f(u,\chi,\Psi)$, where the only constraint on $f$ is
$\partial_{\Psi}f\neq 0$\footnote{As this will in general change
${E^{\bullet}}_u$, one must compensate by a local Lorentz transformation
\eqref{transf-E} in order to preserve the gauge condition ${E^{\bullet}}_u=0$.}.
Choosing $f=\Xi$ we have then
\eq
\texttt{A}_u = \texttt{A}_{\chi} = 0\ , \qquad \texttt{A}_{\Psi} =
(\X\,\Xb\tanh\chi)^{1/2}\ ,
\feq
and thus ${E^+}_u={E^+}_{\chi}={E^+}_{\Psi}=0$. This means that the most
general half-supersymmetric background in this class is given by
\eq
ds^2 = -2\sqrt2\coth\chi dudv + \frac{d\chi^2}{16g^2\sinh^2\!\chi \X\,\Xb}
+ \frac{d\Psi^2}{2\sinh\!2\chi}\ ,
\feq
\eq
F^I = \frac{(\mbox{Im}\,\mathcal{N})^{-1|IJ}\xi_J}{4\cosh^2\!\chi
(\X\,\Xb\tanh\chi)^{1/2}}\,d\Psi\wedge d\chi\ ,
\feq
while the scalars $z^{\alpha}(\chi)$ follow from the flow
equation \eqref{z-chi-Psi}. Hence, all the solutions of
section \ref{expl-dchineq0} with ${\cal G}=0$ are actually half-BPS.

Integration of the Killing spinor equations \eqref{u1}, \eqref{u2},
\eqref{achi}, \eqref{aPsi} and \eqref{bchi} yields
\eq
b = b_0\left(\frac{\X}{\Xb\sinh\!2\chi}\right)^{1/2}\ ,
\qquad a = a_0\ ,
\feq
where $a_0,b_0$ are constants. In what follows, we shall take
$b_0=1$ without loss of generality (for $b_0=0$ one gets the first Killing
spinor). Then the functions $c$ and $d$ read
\eq
c = e^{\chi}{\bar a}_0\ , \qquad d = e^{-\chi}b\ .
\feq
The bilinear $V_{\mu}=A(\epsilon^i,\Gamma_{\mu}\epsilon_i)$ associated
to the second covariantly constant spinor has norm squared
\eq
V^2 = -8\frac{(\mbox{Re}a_0\sinh\chi)^2+(\mbox{Im}a_0\cosh\chi)^2}
      {|\sinh\chi|\cosh\chi}\ ,
\feq
which is in general negative, so that the solution belongs also to the timelike
class studied in \cite{Cacciatori:2008ek}.
For $a_0=0$ and $\chi>0$, the second Killing vector is given by $V=\partial_u$.

Note that the $uu$ component of the Einstein equations is identically
satisfied for the half-supersymmetric backgrounds. This is not surprising,
since they belong also to the timelike class, where the Killing spinor
equations imply all the equations of motion \cite{Cacciatori:2008ek}.

\subsection{Case $d\chi=0$}
\label{1/2-dchi0}

From the vanishing of the gravitino variation we obtain
the system
\begin{eqnarray}
\partial_u a &=& \frac{i\psi}{\sqrt2H}(a-\bar c) - \frac{{E^+}_{u,w}-{E^+}_{w,u}}
{{E^\bullet}_w}b + ig\sqrt2{E^+}_u(b{\bar X}^I-\bar d X^I)\xi_I\ ,
\nonumber \\
\partial_u b &=& -iA_ub + \frac{i\psi}{\sqrt2H}b + \frac{ig\sqrt2}{H}\X
(a-\bar c)\ , \nonumber \\
\partial_u c &=& \frac{i\psi}{\sqrt2H}(\bar a-c) - \frac{{E^+}_{u,w}-{E^+}_{w,u}}
{{E^\bullet}_w}d - ig\sqrt2{E^+}_u(\bar b X^I-d{\bar X}^I)\xi_I\ , \nonumber \\
\partial_u d &=& -iA_ud - \frac{i\psi}{\sqrt2H}d - \frac{ig\sqrt2}{H}\X
(\bar a-c)\ , \label{abl_u}
\end{eqnarray}
\begin{eqnarray}
\partial_v a &=& -4ig(b{\bar X}^I-\bar dX^I)\xi_I\ , \nonumber \\
\partial_v b &=& \partial_v d = 0\ , \nonumber \\
\partial_v c &=& 4ig(\bar bX^I-d{\bar X}^I)\xi_I\ , \label{abl_v}
\end{eqnarray}
\begin{eqnarray}
\partial_w a &=& -\frac{i\psi}{\sqrt2}{E^\bullet}_w\bar d + ig\sqrt2{E^+}_w
(b{\bar X}^I-\bar dX^I)\xi_I\ , \nonumber \\
\partial_w b &=& -iA_w b - ig\sqrt2\X{E^\bullet}_w\bar d\ , \nonumber \\
\partial_w c &=& \frac{i\psi}{\sqrt2}{E^\bullet}_w\bar b - ig\sqrt2{E^+}_w
(\bar bX^I-d{\bar X}^I)\xi_I\ , \nonumber \\
\partial_w d &=& -iA_w d - ig\sqrt2\X{E^\bullet}_w\bar b\ , \label{abl_w}
\end{eqnarray}
\begin{eqnarray}
\partial_{\bar w} a &=& -\left(iA_u-\partial_u\ln{E^{\bar\bullet}}_{\bar w}\right)
{E^{\bar\bullet}}_{\bar w}Hb - ig\sqrt2\X{E^{\bar\bullet}}_{\bar w}(a-\bar c)
+ ig\sqrt2{E^+}_{\bar w}(b{\bar X}^I-\bar dX^I)\xi_I\ , \nonumber \\
\partial_{\bar w}b &=& -\left(iA_{\bar w}-ig\sqrt2\X{E^{\bar\bullet}}_{\bar w}
\right)b\ , \nonumber \\
\partial_{\bar w} c &=& -\left(iA_u-\partial_u\ln{E^{\bar\bullet}}_{\bar w}\right)
{E^{\bar\bullet}}_{\bar w}Hd + ig\sqrt2\X{E^{\bar\bullet}}_{\bar w}(\bar a-c)
- ig\sqrt2{E^+}_{\bar w}(\bar bX^I-d{\bar X}^I)\xi_I\ , \nonumber \\
\partial_{\bar w}d &=& -\left(iA_{\bar w}-ig\sqrt2\X{E^{\bar\bullet}}_{\bar w}
\right)d\ , \label{abl_wquer}
\end{eqnarray}
while the gaugino supersymmetry transformations yield
\begin{eqnarray}
\partial_u z^{\alpha} &=& \frac{ig\sqrt2}{2H}e^{{\cal K}/2}g^{\alpha\bar\beta}
{\cal D}_{\bar\beta}{\bar Z}^I\xi_I\left(\frac{\bar a-c}{\bar b}-\frac{a-\bar c}
{\bar d}\right)\ , \label{duz} \\
\partial_{\bar w} z^{\alpha} &=& -ig\sqrt2 e^{{\cal K}/2}g^{\alpha\bar\beta}
{\cal D}_{\bar\beta}{\bar Z}^I\xi_I\frac b{\bar d}{E^{\bar\bullet}}_{\bar w}\ ,
\label{dbarwz} \\
0 &=& g^{\alpha\bar\beta}{\cal D}_{\bar\beta}{\bar Z}^I\left[ig\xi_I
\left(\frac{\bar a-c}{\bar b}+\frac{a-\bar c}{\bar d}\right) - 2(\mbox{Im}\,
\mathcal{N})_{IJ}\psi^J\frac b{\bar d}\right]\ , \label{gaugini3}
\end{eqnarray}
as well as $\bar b b=\bar d d$. Note that we assume that both $b$ and $d$ are
nonvanishing, because for $b=0$ or $d=0$ the only solution to
\eqref{abl_u}-\eqref{abl_wquer} is the first Killing spinor.
From \eqref{dwz}, \eqref{duz} and \eqref{dbarwz} one gets
\begin{eqnarray}
\partial_u(\X\,\Xb) &=& \frac{ig}{\sqrt2 H}g^{\alpha\bar\beta}{\cal D}_{\alpha}X^I{\cal D}_{\bar\beta}{\bar X}^J
\xi_I\xi_J\left(\Xb+\X\frac{\bar d}b\right)\left(\frac{\bar a-c}{\bar b}-\frac{a-\bar c}{\bar d}\right)\ , \nonumber \\
\partial_w(\X\,\Xb) &=& ig\sqrt2{E^\bullet}_wg^{\alpha\bar\beta}{\cal D}_{\alpha}X^I{\cal D}_{\bar\beta}
{\bar X}^J\xi_I\xi_J\left(\Xb+\X\frac{\bar d}b\right)\ , \label{dwX}
\end{eqnarray}
that will be useful later. Equ.~\eqref{gaugini3} allows to determine $\psi^I$,
\eq
\psi^I = i\psi{\bar X}^I-ig\left(\frac{a-\bar c}b+\frac{\bar a-c}d\right)
g^{\alpha\bar\beta}{\cal D}_{\alpha}X^I{\cal D}_{\bar\beta}{\bar X}^J\xi_J\ .
\label{gaugini3'}
\feq
The $u$-$v$, $w$-$v$ and $\bar w$-$v$ integrability conditions imply
\begin{eqnarray}
b(\partial_\mu-iA_\mu)\Xb - \bar d(\partial_\mu+iA_\mu)\X &=& 0\ , \label{DmuX} \\
\psi\left({\bar X}^Ib-X^I\bar d\right)\xi_I &=& 0\ . \label{bd}
\end{eqnarray}
From \eqref{bd} we have $\psi=0$ or
\eq
d = \frac{\X}{\Xb}\bar b\ . \label{bd-final}
\feq
Let us first consider the latter case \eqref{bd-final}. Then, \eqref{DmuX} gives
\eq
A_{\mu}=\frac i2\partial_{\mu}\ln\left(\frac{\X}{\Xb}\right)\ .
\label{Amu}
\feq
Notice that this follows also from \eqref{dwz}, \eqref{duz}, \eqref{dbarwz} and \eqref{bd-final}.

Using \eqref{Amu}, equ.~\eqref{FI4} can be readily integrated, with the result
\eq
{E^\bullet}_w = H^{-1/2}\left(\frac{\X}{\Xb}\right)^{1/2}f(u,w)\ ,
\feq
where $f(u,w)$ denotes an arbitrary function that can be set to one without
loss of generality by a reasoning analogous to that following \eqref{f(u,w)}.
Plugging \eqref{Amu} into \eqref{FI1} leads to ${E^+}_w=\partial_w m$, with
$m$ some real function. By shifting $v$ and $\cal G$ appropriately, one can
always achieve $m=0$. \eqref{FI2} simplifies to
\eq
\partial_w\sqrt H = \sqrt2ig(\X\,\Xb)^{1/2}\ ,
\label{FI2'}
\feq
which implies
\eq
(\partial_w+\partial_{\bar w})H = 0\ ,
\feq
so that $H$ depends on $w-\bar w$ and $u$ only. Moreover, combining \eqref{dbarwz}
with the flow equation \eqref{dwz}, we obtain
\eq
(\partial_w+\partial_{\bar w})z^{\alpha} = 0\ , \label{z=z(u,y)}
\feq
and thus the scalars are independent of $w+\bar w$ as well. The remaining
integrability conditions for the system \eqref{abl_u}-\eqref{abl_wquer} turn
out to be
\begin{eqnarray}
0 &=& \left[\left(\frac{\Xb}{\X}\right)^{1/2}\sqrt H\partial_u
(\X\,\Xb)+\left(\psi^I-i\psi{\bar X}^I\right)\xi_I
\left(\frac{2\X\,\Xb}H\right)^{1/2}\right]b \nonumber \\
&+&\partial_w(\X\,\Xb)(a-\bar c)\ , \label{1a-int} \\
0 &=& -\frac{ig}{H^2}\frac{\X}{\Xb}\left[\psi^I+2i\psi{\bar X}^I
\right]\xi_I(a-\bar c) + \left[\partial^2_w{\cal G}-\frac{i\partial_u\psi}
{\sqrt2H}\right]b\ , \label{2a-int} \\
0 &=& \frac{ig}{H^{3/2}}\left[\sqrt2H\partial_u(\X\,\Xb)^{1/2}
-\left(\psi^I-i\psi{\bar X}^I\right)\xi_I\left(\frac{\X}
{\Xb}\right)^{1/2}\right](a-\bar c) \label{3a-int} \\
&-&\left(\frac{\Xb}{\X}\right)^{1/2}\left[\frac{\psi^2}{2H^{3/2}}
+\sqrt H\partial_w\partial_{\bar w}{\cal G}-\partial^2_u\sqrt H+ig
(2\X\,\Xb)^{1/2}(\partial_w-\partial_{\bar w}){\cal G}
\right]b\ , \nonumber
\end{eqnarray}
together with
\eq
\left({\bar\psi}^I{\bar X}^J+\psi^IX^J\right)\xi_I\xi_J = 0\ , \label{psiIquer}
\feq
and
\eq
\left(\partial^2_w-\partial^2_{\bar w}\right){\cal G} = 0\ . \label{dxdyG}
\feq
Making use of \eqref{psiIquer} in the complex conjugate of \eqref{F=dAchi=0}
(recall that $\psi$ is real) yields
\eq
(\partial_w+\partial_{\bar w})\frac{\psi}H = 0\ ,
\feq
hence $\psi=\psi(w-\bar w,u)$.
Plugging the eqns. \eqref{dwX} as well as the contraction of \eqref{gaugini3'} with
$\xi_I$ into \eqref{1a-int}, one finds that the latter is identically satisfied.
From \eqref{gaugini3'} and \eqref{2a-int} one obtains
\eq
\frac{a-\bar c}b\partial_w\psi = \left(\frac{\Xb}{\X}\right)^{1/2}
\left[\frac{iH^2\partial_w\psi\,\partial_w(\psi H^{-3/2})}{2\sqrt2g^2
g^{\alpha\bar\beta}{\cal D}_{\alpha}X^I{\cal D}_{\bar\beta}{\bar X}^J\xi_I\xi_J}-
\sqrt H\partial_u\psi\right]\ , \label{a-barc}
\feq
and the constraint
\eq
\partial^2_w{\cal G} = -\frac{H^{1/2}\partial_w\psi\,\partial_w(\psi H^{-3/2})}
{4g^2g^{\alpha\bar\beta}{\cal D}_{\alpha}X^I{\cal D}_{\bar\beta}{\bar X}^J\xi_I\xi_J}\ ,
\label{d2wG}
\feq
where we used
\begin{eqnarray}
\left(\psi^I-i\psi{\bar X}^I\right)\xi_I &=& \left(\frac{\Xb}{\X}
\right)^{1/2}\frac{H^2}{g\sqrt2}\partial_w(\psi H^{-3/2})\ , \nonumber \\
\left(\psi^I+2i\psi{\bar X}^I\right)\xi_I &=& \left(\frac{H\Xb}
{2\X}\right)^{1/2}\frac{\partial_w\psi}g\ , \nonumber
\end{eqnarray}
that follow from \eqref{F=dAchi=0} and \eqref{FI2'}.
Note that above we assumed that $g^{\alpha\bar\beta}{\cal D}_{\alpha}X^I
{\cal D}_{\bar\beta}{\bar X}^J\xi_I\xi_J$ is nonvanishing. If this expression
were zero, then $\xi_I{\cal D}_{\alpha}X^I=0$, since the K\"ahler metric
$g_{\alpha\bar\beta}$ is non-degenerate. As a consequence, the scalars are constant,
as can be seen from \eqref{abl_V}. This case was considered in section
\ref{const-scal} and will not be pursued further here.

In addition to \eqref{dxdyG}and \eqref{d2wG}, the function $\cal G$
must obey the $uu$ component of the Einstein equations, that becomes
\begin{eqnarray}
0 &=& \partial_w\partial_{\bar w}{\cal G} - \frac{\partial^2_u\sqrt H}{\sqrt H}
+ig\sqrt{\frac{2\X\,\Xb}H}(\partial_w-\partial_{\bar w})
{\cal G} \nonumber \\
&& +\frac{ig}{\sqrt{2\X\,\Xb H}}\frac{\left[\partial_u(\X\,\Xb)\right]^2}
{\partial_w(\X\,\Xb)} - \frac{(\mbox{Im}\,\mathcal{N})_{IJ}{\bar\psi}^I\psi^J}
{H^2}\ . \label{siklos}
\end{eqnarray}
Using
\eq
\psi^I = \left(\frac{\Xb}{\X}\right)^{1/2}\frac{H^2\partial_w(\psi H^{-3/2})
g^{\gamma\bar\delta}{\cal D}_{\gamma}X^I{\cal D}_{\bar\delta}{\bar X}^J\xi_J}
{g\sqrt2g^{\alpha\bar\beta}{\cal D}_{\alpha}X^K{\cal D}_{\bar\beta}{\bar X}^L
\xi_K\xi_L}+i\psi{\bar X}^I\ , \label{psiI-fin}
\feq
following from \eqref{gaugini3'} and \eqref{a-barc}, one finds that
\eqref{siklos} is equivalent to the last integrability condition \eqref{3a-int}.

Finally we come to the Maxwell equations and Bianchi identities. As is clear from
\eqref{psiI-fin}, the expression $\psi^I\rho e^{i\zeta}/H$ entering
\eqref{Bianchichi=0} and \eqref{Maxwchi=0} depends on $w-\bar w$ and $u$ only,
such that $(\partial_w+\partial_{\bar w})(\psi^I\rho e^{i\zeta}/H)=0$.
Plugging this into \eqref{Bianchichi=0}, one gets
\eq
\psi^I\rho e^{i\zeta}/H + {\bar\psi}^I\rho e^{-i\zeta}/H = 2l^I(u)\ , \label{l^I}
\feq
where $l^I(u)$ are arbitrary real functions of $u$ obeying the constraint
$l^I\xi_I=0$ due to \eqref{psiIquer}, and the factor $2$ was
chosen for later convenience. By virtue of \eqref{z=z(u,y)}, ${\cal N}_{IJ}$
is likewise independent of $w+\bar w$, and hence \eqref{Maxwchi=0} implies
\eq
{\bar{\cal N}}_{IJ}\psi^J\rho e^{i\zeta}/H + {\cal N}_{IJ}{\bar\psi}^J\rho
e^{-i\zeta}/H  = 2m_I(u)\ , \label{m_I}
\feq
with $m_I(u)$ again some real functions. Since $(\mbox{Im}\,\mathcal{N})_{IJ}$
is invertible, \eqref{l^I} together with \eqref{m_I} give an expression for
$\psi^I$ in terms of $l^I$ and $m_I$,
\eq
\psi^I\rho e^{i\zeta}/H = l^I(u) + i(\mbox{Im}\,\mathcal{N})^{-1|IJ}\left(m_J(u)
-(\mbox{Re}\,\mathcal{N})_{JK}l^K(u)\right)\ .
\feq
Reality of $\psi$ yields in addition
\eq
\Xb\left(F_Il^I-m_IX^I\right)=\X\left({\bar F}_Il^I-m_I{\bar X}^I\right)\ .
\label{psi-real}
\feq
In what follows, we shall solve the above equations for the simple model with
prepotential $F=-iZ^0Z^1$ introduced in section \ref{simple-prepot}.
In this case one has
\eq
\X = \frac{\xi_0+\xi_1z}{\sqrt{2(z+\bar z)}}\ .
\feq
A sufficient condition for \eqref{psi-real} to be satisfied is $l^I=0$ and
$z=\bar z$. Then also the constraint $l^I\xi_I=0$ is met, and we obtain
\eq
\psi^0 = -\frac{im_0}zH^{3/2}\ , \qquad \psi^1 = -im_1zH^{3/2}\ , \qquad
\psi = -H^{3/2}\frac{m_1z+m_0}{\sqrt z}\ .
\feq
\eqref{dwz} and \eqref{FI2'} reduce respectively to
\eq
\partial_w z = ig\left(\frac{2z}H\right)^{1/2}(-\xi_0+\xi_1z)\ ,
\qquad\partial_w\sqrt H = ig\frac{\xi_0+\xi_1z}{\sqrt{2z}}\ . \label{dwzH}
\feq
Making use of this, one finds that \eqref{F=dAchi=0} holds identically.
Moreover, the eqns.~\eqref{dwzH} imply that $H$ is again given by \eqref{H(z)}, where
we choose $f(u)=1$. Let us further assume that the $m_I$ are constants and
$z$ is independent of $u$. $z$ is then a function of $y$ only, where we defined
$w=x+iy$. With these choices, the conditions \eqref{psiI-fin} are fulfilled as well.
To determine the wave profile ${\cal G}$, one first observes that
\eqref{dxdyG} leads to
\eq
{\cal G} = {\cal G}_1(u,x) + {\cal G}_2(u,y)\ .
\feq
From \eqref{d2wG} it is clear that $\partial^2_w{\cal G}$ does not depend on
$x$, and thus
\eq
{\cal G}_1(u,x) = n(u)x^2 + h(u)x + j(u)\ ,
\feq
for some functions $n(u),h(u),j(u)$. By shifting the coordinate $v$, we can always
set $j=0$ without loss of generality. Integrating once the Einstein equation \eqref{siklos}
yields
\eq
\partial_y{\cal G}_2 = H\left[\frac{\sqrt2}g\left(-\frac{m_0^2}z+m_1^2z\right)-\frac{n(u)}{\sqrt2g\xi_1(-\xi_0
+\xi_1z)}+k(u)\right]\ , \label{dyG}
\feq
with $k(u)$ arbitrary at this stage. Compatibility of this with \eqref{d2wG} requires
\eq
k(u) = \frac{\sqrt2m_1}{g\xi_1}(\xi_0m_1-\xi_1m_0)\ , \qquad n(u) = -2(\xi_0m_1-\xi_1m_0)^2\ .
\feq
Before integrating \eqref{dyG} again,  we explicitely solve the Killing spinor equations
\eqref{abl_u}-\eqref{abl_wquer}. From the equations for $\partial_wb$ and $\partial_{\bar w}b$
one obtains $b=\beta(u)/\sqrt H$, where $\beta(u)$ denotes an arbitrary function. Then, \eqref{a-barc} gives
\eq
a-\bar c = \frac1{gz}(-\xi_0+\xi_1z)(m_1z-m_0)\beta(u)\ ,
\feq
Deriving this with respect to $u$ and using the relations \eqref{abl_u}, one gets
\eq
\frac{\beta'(u)}{\beta(u)} = -\frac{in(u)}{\sqrt2\xi_1} \label{beta'}\ ,
\feq
together with $k(u)=0$, i.e.,
\eq
m_1=0 \qquad \text{or} \qquad \xi_0m_1=\xi_1m_0\ .
\feq
We shall choose here the latter possibility. Then $n(u)=0$, and \eqref{beta'} implies that $\beta$ must
be constant, $\beta=\beta_0$. The remaining Killing spinor equations can then be easily integrated,
with the result
\begin{eqnarray}
a &=& \frac{\beta_0}{2g}\left(m_1\xi_1z + \frac{m_0\xi_0}z\right) + \alpha(u)\ , \nonumber \\
c &=& -\frac{{\bar\beta}_0}{2g}\left(m_1\xi_1z + \frac{m_0\xi_0}z\right) + {\bar\alpha}(u)
+ \frac{2{\bar\beta}_0}g\xi_0m_1\ , \nonumber \\
b &=& \bar d = \frac{\beta_0}{\sqrt H}\ , \label{2ndKill}
\end{eqnarray}
where $\alpha(u)$ obeys
\eq
\alpha'(u) = -\frac{\beta_0}2h(u)\ .
\feq
The first Killing spinor is recovered for $\beta_0=0$. The Killing vector associated to \eqref{2ndKill}
has components
\begin{eqnarray}
V_+ &=& 2\sqrt2\frac{|\beta_0|^2}H\ , \nonumber \\
V_- &=& -\frac{|\beta_0|^2}{\sqrt2g^2}\left(m_1\xi_1z+\frac{m_0\xi_0}z-2\xi_0m_1\right)^2
-2\sqrt2|\alpha+\frac{\beta_0}g\xi_0m_1|^2\ , \nonumber \\
V_{\bullet} &=& V_{\bar\bullet} = -\sqrt{\frac2H}\left(\alpha{\bar\beta}_0+{\bar\alpha}\beta_0+
\frac{2|\beta_0|^2}g\xi_0m_1\right)\ ,
\end{eqnarray}
and norm squared
\eq
V^2 = -\frac{4|\beta_0|^4}{g^2}z\left(m_1-\frac{m_0}z\right)^2-\frac{16}H\mbox{Im}^2\left[{\bar\beta}_0
\left(\alpha+\frac{\beta_0}g\xi_0m_1\right)\right]\ ,
\feq
which is in general negative, unless $\beta_0=0$, so that the solution belongs to the timelike class
as well. This explains also why the $uu$ component of the Einstein equations is implied by the
integrability conditions \eqref{3a-int}.

Finally, \eqref{dyG} yields the wave profile
\eq
{\cal G} = {\cal G}_1 + {\cal G}_2 = h(u)x - \frac{m_1^2}{4\xi_1^2g^2}H^2\ ,
\feq
where
\eq
H = 4\xi_0\xi_1\sinh^2\!\sqrt2gy\ .
\feq
The scalar field $z$ and metric are given by \eqref{z(y)} and \eqref{domainwall} respectively,
and the fluxes read
\eq
F^0 = \frac{2m_0}zdu\wedge dy\ , \qquad F^1 = 2m_1z du\wedge dy\ .
\feq
Note that for $y\to0$, when the scalar goes to its critical value,
we have $H\to8\xi_0\xi_1g^2y^2$, and thus (if we choose $h(u)=0$) the wave
profile becomes
\eq
{\cal G} = -16m_1^2\xi_0^2g^2y^4\ ,
\feq
which means that the solution reduces to a subclass of the charged generalization
of the Kaigorodov spacetime found in \cite{Cai:2003xv}.

This concludes the explicit example of a half-supersymmetric background with $d\chi=0$, $\psi\neq0$.\\

Next we have to consider the case when $\psi=0$. Then, \eqref{xiAchi=0} and \eqref{F=dAchi=0} give
\eq
\psi^I\xi_I = A^I\xi_I = F^I\xi_I = 0\ .
\feq
Contracting \eqref{gaugini3'} with $\xi_I$, taking into account that $\psi=\psi^I\xi_I=0$ and assuming as before that $g^{\alpha\bar\beta}{\cal D}_{\alpha}X^I{\cal D}_{\bar\beta}{\bar X}^J\xi_I\xi_J\neq0$ (otherwise,
as was explained above, the scalar fields would be constant), we get
\eq
\frac{a-\bar c}b+\frac{\bar a-c}d=0\ . \label{ac-imag}
\feq
Plugging this back into \eqref{gaugini3'}, one obtains $\psi^I=0$, so there are no fluxes turned on
in this case. The Killing spinor equations together with the integrability conditions \eqref{DmuX}
imply
\eq
\partial_{\mu}(\bar bX^I-d{\bar X}^I)\xi_I=0\ ,
\feq
hence
\eq
(\bar bX^I-d{\bar X}^I)\xi_I=\lambda\ , \label{def-lambda}
\feq
with $\lambda$ a constant.

Let us first assume that $b\Xb + {\bar d}\X=0$, so that $\X\,\Xb$ is constant due to \eqref{dwX}. Then, the
Killing spinor equations for $b$ and $d$ simplify to
\begin{eqnarray}
\partial_u b &=& -iA_ub + \frac{ig\sqrt2}H\X(a-\bar c) = iA_ub - \frac{ig\sqrt2}H\X(a-\bar c) - b\partial_u
\ln\frac{\Xb}{\X}\ , \nonumber \\
\partial_w b &=& -iA_wb + ig\sqrt2\Xb{E^{\bullet}}_wb = iA_wb - ig\sqrt2\Xb{E^{\bullet}}_wb -
b\partial_w\ln\frac{\Xb}{\X}\ , \nonumber \\
\partial_{\bar w} b &=& -iA_{\bar w}b + ig\sqrt2\X{E^{\bar\bullet}}_{\bar w}b = iA_{\bar w}b -
ig\sqrt2\X{E^{\bar\bullet}}_{\bar w}b - b\partial_{\bar w}\ln\frac{\Xb}{\X}\ , \nonumber
\end{eqnarray}
from which one obtains
\eq
A_u = \frac i2\partial_u\ln\frac{\X}{\Xb} + \frac{g\sqrt2}H\X\frac{a-\bar c}b\ , \label{A_u}
\feq
\eq
A_w = \frac i2\partial_w\ln\frac{\X}{\Xb} + g\sqrt2\Xb{E^\bullet}_w\ , \qquad
A_{\bar w} = \frac i2\partial_{\bar w}\ln\frac{\X}{\Xb} + g\sqrt2\X{E^{\bar\bullet}}_{\bar w}\ , \label{A_barw}
\feq
as well as
\[
b = \lambda_1\sqrt{\frac{\X}{\Xb}}\ , \qquad d = -\bar{\lambda}_1\sqrt{\frac{\X}{\Xb}}\ ,
\]
where $\lambda_1\neq 0$ is an integration constant. Using the expression \eqref{A_barw} for $A_{\bar w}$
in \eqref{FI4} leads to
\eq
{E^\bullet}_w = \sqrt{\frac{\X}{\Xb}}f(u,w)\ ,
\feq
with $f(u,w)$ an arbitrary function that, as before, can be set to unity without loss of generality.
Then we have
\[
\rho=1\ , \qquad e^{i\zeta}=\sqrt{\frac{\X}{\Xb}}\ ,
\]
and thus
\[
A_u+\partial_u\zeta = \frac{g\sqrt2}{\lambda_1H}\sqrt{\X\,\Xb}(a-\bar c)\ ,
\]
where we used \eqref{A_u}. This, together with \eqref{FI1} gives the relation 
\eq
c = \bar a + \frac{i{\bar\lambda}_1}{2g\sqrt{2\X\,\Xb}}({E^+}_{\bar w,w}-{E^+}_{w,\bar w}) \label{caquer}
\feq
between $\bar a$ and $c$, that can be substituted into the Killing spinor equations for $a$ and $c$,
which become
\begin{eqnarray}
\frac{\partial_u a}{\lambda_1} &=& 2ig\sqrt{2\X\,\Xb}{E^+}_u - ({E^+}_{u,w}-{E^+}_{w,u}) \nonumber \\
&=& 2ig\sqrt{2\X\Xb}{E^+}_u + ({E^+}_{u,\bar w}-{E^+}_{\bar w,u}) +
\frac i{2g\sqrt{2\X\,\Xb}}\partial_u({E^+}_{w,\bar w}-{E^+}_{\bar w,w})\ , \nonumber \\
\frac{\partial_v a}{\lambda_1} &=& -8ig\sqrt{\X\,\Xb}\ , \nonumber \\
\frac{\partial_w a}{\lambda_1} &=& 2ig\sqrt{2\X\,\Xb}{E^+}_w \nonumber \\
&=& 2ig\sqrt{2\X\Xb}{E^+}_w + ({E^+}_{w,\bar w}-{E^+}_{\bar w,w}) +
\frac i{2g\sqrt{2\X\,\Xb}}\partial_w({E^+}_{w,\bar w}-{E^+}_{\bar w,w})\ , \nonumber \\
\frac{\partial_{\bar w} a}{\lambda_1} &=& 2ig\sqrt{2\X\Xb}{E^+}_{\bar w} +
\frac i{2g\sqrt{2\X\,\Xb}}\partial_{\bar w}({E^+}_{w,\bar w}-{E^+}_{\bar w,w}) \nonumber \\
&=& 2ig\sqrt{2\X\,\Xb}{E^+}_{\bar w} + ({E^+}_{w,\bar w}-{E^+}_{\bar w,w})\ , \label{KSE-final-a}
\end{eqnarray}
so that $E^+$ is constrained by
\begin{eqnarray}
\partial_u({E^+}_{\bar w,w}-{E^+}_{w,\bar w}) &=& 2ig\sqrt{2\X\,\Xb}({E^+}_{w,u}-{E^+}_{u,w}+
{E^+}_{\bar w,u}-{E^+}_{u,\bar w})\ , \label{1stconstrE^+} \\
\partial_w({E^+}_{\bar w,w}-{E^+}_{w,\bar w}) &=& 2ig\sqrt{2\X\,\Xb}({E^+}_{\bar w,w}-{E^+}_{w,\bar w})\ , \nonumber \\
\partial_{\bar w}({E^+}_{\bar w,w}-{E^+}_{w,\bar w}) &=& -2ig\sqrt{2\X\,\Xb}({E^+}_{\bar w,w}-{E^+}_{w,\bar w})\ . \nonumber
\end{eqnarray}
Integrating the last two equations, one has
\eq
{E^+}_{\bar w,w}-{E^+}_{w,\bar w} = {\cal E}(u)\exp\left[2ig\sqrt{2\X\,\Xb}(w-\bar w)\right]\ ,
\feq
with ${\cal E}(u)$ some imaginary function. This implies
\eq
{E^+}_w - \frac{{\cal E}(u)}{4ig\sqrt{2\X\,\Xb}}\exp\left[2ig\sqrt{2\X\,\Xb}(w-\bar w)\right] = \partial_w m
\feq
for some real function $m$. By shifting $v$ and $\cal G$ appropriately, we can thus set
${E^+}_w={E^+}_{\bar w}=0$. Then, \eqref{1stconstrE^+} gives
\eq
(\partial_w+\partial_{\bar w}){E^+}_u = 0\ ,
\feq
while from \eqref{caquer} one obtains $c=\bar a$. The Killing spinor equations \eqref{KSE-final-a} are
easily integrated, with the result
\eq
a = -8ig\sqrt{\X\,\Xb}\lambda_1v + \lambda_1\alpha(u)\ ,
\feq
where $\alpha$ satisfies
\eq
\alpha'(u) = 2ig\sqrt{2\X\,\Xb}{E^+}_u - \partial_w{E^+}_u\ . \label{alpha'}
\feq
The latter relation determines ${E^+}_u$,
\eq
{E^+}_u = \frac{\alpha'(u)}{2ig\sqrt{2\X\,\Xb}}+\hat{{\cal E}}(u)\exp\left[2ig\sqrt{2\X\,\Xb}(w-\bar w)\right]\ ,
\feq
with $\hat{\cal E}(u)$ real and otherwise arbitrary. Note that ${E^+}_u$ is independent of $w+\bar w$.
Eqns.~\eqref{dwz}, \eqref{duz} and \eqref{dbarwz} boil down to $\partial_u z^{\alpha} =0$ and
\eq
\partial_w z^{\alpha}  = \partial_{\bar w} z^{\alpha} = ig\sqrt2g^{\alpha\bar\beta}{\cal D}_{\bar\beta}
{\bar X}^I\xi_I\sqrt{\frac{\X}{\Xb}}\ , \label{dxz}
\feq
so that the scalar fields are functions of $x=(w+\bar w)/2$ only. The $uu$ component of the Einstein
equations reads
\[
\partial_w\partial_{\bar w}{E^+}_u + ig\sqrt{2\X\,\Xb}(\partial_w-\partial_{\bar w}){E^+}_u
= 2g^2\left[(\mbox{Im}\,\mathcal{N})^{-1|IJ}\xi_I\xi_J+4\X\,\Xb\right]{E^+}_u\ .
\]
While the lhs is independent of $x$, the prefactor of ${E^+}_u$ on the rhs depends in general
nontrivially on $x$. This is compatible only if ${E^+}_u=0$, hence $\alpha(u)$ is constant due
to \eqref{alpha'}. The function $H$ appearing in the metric follows from \eqref{FI2}, yielding
\eq
H = h(u)\exp\left[-4g\sqrt{2\X\,\Xb}y\right]\ ,
\feq
where $y=(w-\bar w)/2i$, and we can always choose $h(u)=-2\sqrt2$ by redefining the coordinate $u$. 

In conclusion, the metric is given by
\eq
ds^2 = 2\left\{\exp\left[4g\sqrt{2\X\,\Xb}y\right]dudv + dy^2 + dx^2\right\}\ ,
\feq
which is simply AdS$_3$$\times\bR$. The dependence of the scalars on the $\bR$-coordinate $x$
is governed by \eqref{dxz}, that can be rewritten as
\eq
\frac{dz^{\alpha}}{dx} = 2ig\sqrt{2{\cal C}}g^{\alpha\bar\beta}\partial_{\bar\beta}\ln(\Xb e^{{\cal K}/2})\ ,
\feq
where the constant $\cal C$ is defined by
\[
{\cal C} = \X\,\Xb\ .
\]
The solution to the Killing spinor equations reads
\eq
a = \bar c = -8ig\sqrt{\X\,\Xb}\lambda_1v + \lambda_1\alpha\ , \qquad b = \lambda_1\sqrt{\frac{\X}{\Xb}}\ ,
\qquad d = -{\bar\lambda}_1\sqrt{\frac{\X}{\Xb}}\ , \label{spin-ads3}
\feq
which reduces to the first covariantly constant spinor if we rescale $\alpha\to\alpha/\lambda_1$ and
then take $\lambda_1\to 0$. The Killing vector constructed from \eqref{spin-ads3} has components
\begin{eqnarray}
V_+ &=& 2\sqrt2|\lambda_1|^2\ , \qquad V_- = -2\sqrt2|-8ig\sqrt{\X\,\Xb}\lambda_1v + \lambda_1\alpha|^2\ ,
\nonumber \\
V_\bullet &=& 2|\lambda_1|^2[16ig\Xb v+\sqrt{\frac{\Xb}{\X}}({\bar\alpha}-\alpha)]\ , \qquad
V_{\bar\bullet} = {\overline{V_\bullet}}\ ,
\end{eqnarray}
and norm squared
\eq
V^2 = -4|\lambda_1|^4({\bar\alpha}+\alpha)^2\ ,
\feq
which is negative unless $\lambda_1=0$ or $\mbox{Re}\,\alpha=0$, so in general the solution
again belongs also to the timelike class.

The final case to consider is $\psi=0$, $b\Xb+{\bar d}\X\neq0$. Then,
equ.~\eqref{DmuX} together with \eqref{def-lambda} implies that
\begin{eqnarray}
A_\mu&=&-i\frac{b\partial_\mu\Xb-\bar d\partial_\mu\X}{b\Xb+\bar d\X}
\label{A-varsigma} \\
&=&\left(1-\frac{\bar\lambda}{2\Xb b}\right)^{-1}\left(i\partial_\mu\ln
\sqrt{\frac{\X}{\Xb}}-i\frac{\bar\lambda}{\Xb b}\partial_\mu\ln\sqrt{\X}\right)\ .
\nonumber
\end{eqnarray}
One easily shows that $\partial_\mu A_\nu-\partial_\nu A_\mu=0$, and thus
$A_\mu=\partial_\mu\varsigma$ for some $v$-independent function $\varsigma$.
Using this and \eqref{FI2}, we can integrate \eqref{FI4} to obtain
\eq
{E^\bullet}_w = e^{-i\alpha}H^{-1/2}f(u,w)\ ,
\feq
where $f(u,w)$ denotes an arbitrary function that, as was explained before,
can be set to unity without loosing generality. Then one has $\rho=H^{-1/2}$
and $\zeta=-\varsigma$, and \eqref{FI1} yields ${E^+}_w={E^+}_{\bar w}=0$.
The Killing spinor equations for $b$ reduce to
\begin{eqnarray}
\label{bu1}\partial_ub&=&-ib\partial_u\varsigma+\frac{ig\sqrt2}H\X
(a-\bar c)\ , \nonumber \\
\label{bw1}\partial_wb&=&-b\partial_w\left(\frac12\ln H+i\varsigma\right)
+\frac{ig{\bar\lambda}\sqrt2}{\sqrt H}e^{-i\varsigma}\ , \nonumber \\
\label{bwbar1}\partial_{\bar w}\ln b&=&-\partial_{\bar w}\left(\frac12\ln H
+i\varsigma\right)\ ,
\end{eqnarray}
from which we get
\begin{equation}
b=(ig{\bar\lambda}\sqrt2w+\hat{b}(u))e^{-i\varsigma}H^{-1/2}\ , \label{b}
\end{equation}
with $\hat{b}$ obeying
\begin{equation}
\label{bcap}\partial_u\hat{b}=ig\sqrt2\X(a-\bar c)e^{i\varsigma}H^{-1/2}
+\frac12(ig{\bar\lambda}\sqrt2w+\hat{b})\partial_u\ln H\ .
\end{equation}
It follows from the Killing spinor equations for $a$ and $c$ that
\begin{equation}
\partial_\mu c=\frac{d}{b}\partial_\mu a\ .
\end{equation}
Since \eqref{def-lambda} combined with $\bar bb=\bar dd$ leads to
\eq
\frac{\lambda}{\bar\lambda} = -\frac db\ , \label{lambda-db}
\feq
one obtains\footnote{Here we assume $\lambda\neq0$. The case $\lambda=0$,
i.e. $\bar b\X=d\Xb$, was already considered earlier.}
\eq
c=-\frac{\lambda}{\bar\lambda}a+\kappa\ ,
\feq
where $\kappa$ is a constant that satisfies $\bar\lambda\kappa=\lambda\bar\kappa$
due to \eqref{ac-imag}. The Killing spinor equations for $a$ boil down to
\begin{eqnarray}
\partial_ua&=&ig\bar\lambda\sqrt2{E^+}_u-(ig\bar\lambda\sqrt2w+\hat{b})
\partial_w{E^+}_u\ , \nonumber \\
\partial_va&=&-4ig\bar\lambda\ , \nonumber \\
\partial_wa&=&0\ , \nonumber \\
\partial_{\bar w}a&=&-\partial_u\hat{b}\ , \nonumber
\end{eqnarray}
so that
\begin{equation}
a=-4ig\bar\lambda v-\bar w\partial_u\hat{b}+\hat{a}(u)\ ,
\end{equation}
where $\hat{a}$ satisfies
\begin{equation}
\label{acap}\partial_u\hat{a}=\bar w\partial_u^2\hat{b}+ig\bar\lambda\sqrt2
{E^+}_u-(ig\bar\lambda\sqrt2w+\hat{b})\partial_w{E^+}_u\ .
\end{equation}
\eqref{bcap} becomes
\begin{eqnarray}
\label{bcapa}\partial_u\hat{b}&=&\frac1{2\lambda}(\lambda\bar w\partial_u\hat{b}
+\bar\lambda w\partial_u\bar{\hat{b}}-\lambda\hat{a}-\bar\lambda\bar{\hat{a}}
+\bar\lambda\kappa)\partial_{\bar w}\ln H \nonumber \\
&&+\frac12(ig\bar\lambda\sqrt2w+\hat{b})\partial_u\ln H\ .\nonumber
\end{eqnarray}
Deriving (\ref{acap}) with respect to $w$ we obtain $\partial^2_w{E^+}_u=0$, hence
\begin{equation}
{E^+}_u=\omega_1(u)w\bar w+\omega_2(u)w+\bar\omega_2(u)\bar w+\omega_3(u)\ ,
\end{equation}
where $\omega_1$ and $\omega_3$ are real. Shifting the coordinate $v$ one can set
$\omega_3=0$ without loss of generality. Plugging back this expression for
${E^+}_u$ into (\ref{acap}) one gets
\eq
\partial_u\hat{a}=-\omega_2\hat{b}\ , \qquad
\partial_u^2\hat{b}=\omega_1\hat{b}-ig\sqrt2\bar\lambda\bar\omega_2\ .
\feq
Note that ${E^+}_u$ must in addition satisfy the $uu$ component of the Einstein
equations, namely
\begin{eqnarray}
\partial_w\partial_{\bar w}{E^+}_u&=&2g\sqrt{\frac2H}\mbox{Im}\left(\X e^{i\varsigma}
\partial_w{E^+}_u\right)+\frac12\partial_u^2\ln H+\frac14\left(\partial_u\ln H
\right)^2 \nonumber \\
&&+\frac{2ig\sqrt2\partial_u(\X\,\Xb)\mbox{Re}\left(\lambda\bar w\partial_u\hat{b}
-\lambda\hat{a}+\frac{\bar\lambda\kappa}2\right)}{2\sqrt H\bar\lambda\X
\left(\bar{\hat{b}}-ig\lambda\sqrt2\bar w\right)e^{i\varsigma}-H|\lambda|^2}\ .
\end{eqnarray}
Solving these equations in general seems to be difficult. A simplification
can be made by assuming $a=\bar c$, which happens for
$\partial_u\hat b=\partial_u\hat a=0$,
$\lambda\hat a+{\bar\lambda}{\bar{\hat a}}=\lambda\bar\kappa$. If we take in
addition $\omega_1=\omega_2=\omega_3=0$ (and thus ${E^+}_u=0$), \eqref{acap}
is satisfied. Moreover, from \eqref{bcap} one gets $\partial_uH=0$, and
\eqref{duz} yields $\partial_uz^{\alpha}=0$. Note that $\hat b$ can be set to
zero by a constant shift of $w$, cf.~\eqref{b}. Using \eqref{lambda-db}, the
equations \eqref{dwz} and \eqref{dbarwz} simplify to
\begin{eqnarray}
\partial_wz^{\alpha} &=& ig\sqrt2\xi_Ie^{{\cal K}/2}{\cal D}_{\bar\beta}{\bar Z}^I
g^{\alpha\bar\beta}e^{-i\varsigma}H^{-1/2}\ , \nonumber \\
\partial_{\bar w}z^{\alpha} &=& -ig\sqrt2\xi_Ie^{{\cal K}/2}{\cal D}_{\bar\beta}
{\bar Z}^Ig^{\alpha\bar\beta}\frac w{\bar w}e^{-i\varsigma}H^{-1/2}\ ,
\end{eqnarray}
which imply
\eq
(w\partial_w+{\bar w}\partial_{\bar w})z^{\alpha}=0\ ,
\feq
i.e., $\partial_rz^{\alpha}=0$, where we introduced polar coordinates $r,\theta$
according to $w=re^{i\theta}$. The scalar fields depend thus on the angular
coordinate $\theta$ only. This, in turn, gives $A_u=A_r=0$ and
$\varsigma=\varsigma(\theta)$. By means of the separation ansatz
$\sqrt H=rh(\theta)$, \eqref{FI2} becomes
\eq
h(\theta)-ih'(\theta) = 2\sqrt2ig\Xb e^{i(\theta-\varsigma)}\ , \label{h}
\feq
and the flow equation reduces to
\eq
\frac{dz^{\alpha}}{d\theta} = -2\sqrt2g\xi_Ie^{{\cal K}/2}{\cal D}_{\bar\beta}
{\bar Z}^Ig^{\alpha\bar\beta}e^{i(\theta-\varsigma)}h^{-1}\ . \label{dthetaz}
\feq
\eqref{A-varsigma}, with $A_{\theta}=\partial_{\theta}\varsigma$, can be rewritten
as
\eq
e^{i\theta}\partial_{\theta}(\Xb e^{-i\varsigma})=e^{-i\theta}\partial_{\theta}
(\X e^{i\varsigma})\ . \label{A_theta}
\feq
Solving \eqref{h} for $\X e^{i\varsigma}$ and plugging the result into
\eqref{A_theta}, one finds that \eqref{A_theta} holds identically.
In conclusion, the unknown functions $z^{\alpha}$, $h$ and $\varsigma$ are
determined by the system of ordinary differential equations \eqref{h} and
\eqref{dthetaz}.
In the following, we shall solve these equations for the SU$(1,1)$/U$(1)$
model with prepotential $F=(Z^1)^3/Z^0$. Choosing $Z^0=1$, $Z^1=-z$, the symplectic
vector reads
\eq
v = \left(\begin{array}{c} 1 \\ -z \\ z^3 \\ 3z^2\end{array}\right)\ .
\feq
The K\"ahler potential and metric are given respectively by
\eq
e^{-\cal K} = 8({\mbox{Im}}z)^3\ , \qquad g_{z\bar z} = -\frac3{(z-\bar z)^2}\ ,
\feq
so we must have ${\mbox{Im}}z>0$. For the scalar potential one obtains
\eq
V = g^2V_3 = -\frac{4g^2\xi_1^2}{3{\mbox{Im}}z}\ .
\feq
Notice that this model permits to introduce a gauging without having a scalar
potential, by choosing $\xi_1=0$, $\xi_0\neq0$.

Solving \eqref{h} for $e^{i(\theta-\varsigma)}$ and plugging into \eqref{dthetaz}
gives in general
\eq
dz^{\alpha} = ig^{\alpha\bar\beta}\partial_{\bar\beta}\ln(\Xb e^{{\cal K}/2})
(d\theta-id\ln h)\ ,
\feq
which reduces to
\eq
dz = -i(z-\bar z)(d\theta-id\ln h) \label{dz-model}
\feq
for the model under consideration, if we make the choice $\xi_1=0$. Subtracting
this from its complex conjugate yields
\eq
h^2 = \frac A{{\mbox{Im}}z}\ , \label{h^2}
\feq
with $A$ a real positive constant. \eqref{h} implies
\eq
h^2 + {h'}^2 = 8g^2\X\,\Xb\ ,
\feq
which can be easily integrated to give
\eq
{\mbox{Im}}z = \frac{g\xi_0}{\sqrt A}\sin2\theta\ . \label{Imz-theta}
\feq
Positivity of ${\mbox{Im}}z$ restricts $\theta$ to the range $0<\theta<\pi/2$.
Using \eqref{Imz-theta} in the sum of \eqref{dz-model} and its complex conjugate
allows to determine also ${\mbox{Re}}z$. Eventually this leads to
\eq
z = z_0 - \frac{g\xi_0}{\sqrt A}e^{-2i\theta}\ ,
\feq
where $z_0$ denotes a real constant. Then, \eqref{h^2} yields
\eq
h^2(\theta) = \frac{A^{3/2}}{g\xi_0\sin2\theta}\ ,
\feq
so that
\eq
H = \frac{r^2A^{3/2}}{g\xi_0\sin2\theta}\ .
\feq
Finally, \eqref{h} determines $\varsigma=3\theta$.
The metric becomes
\eq
ds^2 = \frac{2g\xi_0\sin2\theta}{A^{3/2}}\left[-\frac{2\sqrt2dudv}{r^2}+\frac{dr^2}
{r^2}+d\theta^2\right]\ ,
\feq
and thus the spacetime is conformal to AdS$_3$ times an interval.

\acknowledgments

This work was partially supported by INFN, MIUR-PRIN contract 20075ATT78 and
by the European Community FP6 program MRTN-CT-2004-005104. We would like to
thank M.~M.~Caldarelli, D.~S.~Mansi, D.~Roest, A.~Van Proeyen and especially
M.~H\"ubscher, P.~Meessen, T.~Ort\'{\i}n and S.~Vaul\`a for useful discussions.
We also wish to thank the Instituto de F\'{\i}sica Te\'orica
UAM/CSIC for hospitality. D.~K.~thanks CERN for hospitality during the
workshop "Black holes: A landscape of theoretical physics problems", where
part of this work was done.

\normalsize

\appendix

\section{Conventions}
\label{conv}

We use the notations and conventions of \cite{Vambroes}, which are briefly summarized here. More
details can be found in appendix A of \cite{Vambroes}.

The signature is mostly plus. Late greek letters $\mu,\nu,\ldots$ are curved spacetime indices,
while early latin letters $a,b,\ldots=0,\ldots,3$ and $A,B,\ldots=+,-,\bullet,\bar\bullet$ refer to the
corresponding tangent space, cf.~also appendix \ref{spin-forms}.

Self-dual and anti-self-dual field strengths are defined by
\begin{equation}
F^{\pm I}_{ab} = \frac 12(F^I_{ab}\pm \tilde F^I_{ab})\,,
\qquad \tilde F^I_{ab} \equiv -\frac i2\epsilon_{abcd}F^{I cd}\,,
\end{equation}
where $\epsilon_{0123}=1$, $\epsilon^{0123}=-1$. We also introduce
\begin{equation}
\epsilon^{\mu\nu\rho\sigma} = e\,e^{\mu}_a e^{\nu}_b e^{\rho}_c e^{\sigma}_d\epsilon^{abcd}\,.
\end{equation}
The $p$-form associated to an antisymmetric tensor $T_{\mu_1\ldots\mu_p}$ is
\begin{equation}
T = \frac 1{p!}T_{\mu_1\ldots\mu_p}dx^{\mu_1}\wedge\ldots\wedge dx^{\mu_p}\,, \label{T}
\end{equation}
and the exterior derivative acts as\footnote{Our definitions for $p$-forms, equ.~\eqref{T}, and for
exterior derivatives, equ.~\eqref{dT}, are the only points where our conventions differ from those
of \cite{Vambroes}.}
\begin{equation}
dT = \frac 1{p!}T_{\mu_1\ldots\mu_p,\nu}dx^{\nu}\wedge dx^{\mu_1}\wedge\ldots\wedge dx^{\mu_p}\,.
\label{dT}
\end{equation}
Antisymmetric tensors are often contracted with $\Gamma$-matrices as in
$\Gamma\cdot F\equiv \Gamma^{ab}F_{ab}$. Moreover, we defined $\X\equiv X^I\xi_I$.

$i,j,\ldots=1,2$ are SU(2) indices, whose raising and lowering is done by complex conjugation.
The Levi-Civita $\epsilon^{ij}$ has the property
\begin{equation}
\epsilon_{ij}\epsilon^{jk} = -{\delta_i}^k\,,
\end{equation}
where in principle $\epsilon^{ij}$ is the complex conjugate of $\epsilon_{ij}$, but we can choose
$\epsilon=i\sigma_2$, such that
\begin{equation}
\epsilon_{12} = \epsilon^{12} = 1\,.
\end{equation}
The Pauli matrices ${\sigma_{xi}}^j$ ($x=1,2,3$) are given by
\begin{equation}
\sigma_1 = \left(\begin{array}{cc} 0 & 1 \\ 1 & 0\end{array}\right)\,, \qquad
\sigma_2 = \left(\begin{array}{cc} 0 & -i \\ i & 0\end{array}\right)\,, \qquad
\sigma_3 = \left(\begin{array}{cc} 1 & 0 \\ 0 & -1\end{array}\right)\,.
\end{equation}
They allow to switch from SU(2) indices to vector quantities using the convention
\begin{equation}
{A_i}^j \equiv i\vec A\cdot\vec\sigma_i^{\;\;j}\,.
\end{equation}
At various places in the main text we use $\sigma$-matrices with only lower or upper indices, defined by
\begin{equation}
\vec\sigma_{ij}\equiv \vec\sigma_i^{\;\;k}\epsilon_{kj}\,, \qquad i\vec\sigma^{ij} = (i\vec\sigma_{ij})^{\ast}\,.
\end{equation}
Notice that both $\vec\sigma_{ij}$ and $\vec\sigma^{ij}$ are symmetric.

Spinors carrying an index $i$ are chiral, e.g. for the supersymmetry parameter one has
\begin{equation}
\Gamma_5\epsilon^i = \epsilon^i\,, \qquad \Gamma_5\epsilon_i = -\epsilon_i\,,
\label{Gamma5-eps}
\end{equation}
and the same holds for the gravitino $\psi^i_{\mu}$. Note however that for some spinors, the upper index
denotes negative chirality rather than positive chirality, for instance the gauginos obey
\begin{equation}
\Gamma_5\lambda^{\alpha i} = -\lambda^{\alpha i}\,, \qquad \Gamma_5\lambda^{\alpha}_i =
\lambda^{\alpha}_i\,,
\end{equation}
as is also evident from the supersymmetry transformations. The charge conjugate of a spinor $\chi$
is
\begin{equation}
\chi^C = \Gamma_0 C^{-1}\chi^{\ast}\,,
\end{equation}
with the charge conjugation matrix $C$. Majorana spinors are defined by $\chi=\chi^C$,
and chiral spinors obey $\chi^C_i=\chi^i$.

\section{Spinors and forms}
\label{spin-forms}

In this appendix, we summarize the essential information needed to realize
the spinors of Spin(3,1) in terms of forms. For more details, we refer
to \cite{Lawson:1998yr}.
Let $V = \bR^{3,1}$ be a real vector space equipped with the Lorentzian inner
product $\langle\cdot,\cdot\rangle$. Introduce an orthonormal basis $e_1, e_2, e_3, e_0$,
where $e_0$ is along the time direction, and consider the subspace $U$
spanned by the first two basis vectors $e_1, e_2$. The space of Dirac spinors
is $\Delta_c = \Lambda^{\ast}(U\otimes \bC)$, with basis
$1, e_1, e_2, e_{12} = e_1 \wedge e_2$.
The gamma matrices are represented on $\Delta_c$ as
\eqn
\Gamma_{0}\eta&=&-e_2\wedge\eta+e_2\rfloor\eta\,, \qquad
\Gamma_{1}\eta=e_1\wedge\eta+e_1\rfloor\eta\,, \nonumber \\
\Gamma_{2}\eta&=&e_2\wedge\eta+e_2\rfloor\eta\,, \qquad
\Gamma_{3}\eta=ie_1\wedge\eta-ie_1\rfloor\eta\,,
\feqn
where
\begin{displaymath}
\eta = \frac 1{k!}\eta_{j_1\ldots j_k} e_{j_1}\wedge\ldots\wedge e_{j_k}
\end{displaymath}
is a $k$-form and
\begin{displaymath}
e_i \rfloor \eta = \frac 1{(k-1)!}\eta_{ij_1\ldots j_{k-1}} e_{j_1}\wedge\ldots
                  \wedge e_{j_{k-1}}\,.
\end{displaymath}
One easily checks that this representation of the gamma matrices satisfies
the Clifford algebra relations $\{\Gamma_a, \Gamma_b\} = 2\eta_{ab}$.
The parity matrix is defined by $\Gamma_5 = i\Gamma_0\Gamma_1\Gamma_2\Gamma_3$,
and one finds that the even forms $1, e_{12}$ have positive chirality,
$\Gamma_5\eta = \eta$, while the odd forms $e_1, e_2$ have negative chirality,
$\Gamma_5\eta = -\eta$, so that $\Delta_c$ decomposes into two complex chiral
Weyl representations $\Delta_c^+ = \Lambda^{\mathrm{even}}(U\otimes \bC)$ and
$\Delta_c^- = \Lambda^{\mathrm{odd}}(U\otimes \bC)$. Note that Spin(3,1) is
isomorphic to SL$(2,\bC)$, which acts with the fundamental representation on the
positive chirality Weyl spinors.\\
Let us define the auxiliary inner product
\begin{equation}
\langle\sum_{i=1}^2 \alpha_i e_i, \sum_{j=1}^2 \beta_j e_j\rangle = \sum_{i=1}^2
\alpha_i^{\ast}\beta_i
\end{equation}
on $U\otimes \bC$, and then extend it to $\Delta_c$. The Spin(3,1) invariant
Dirac inner product is then given by
\begin{equation}
D(\eta, \theta) = \langle\Gamma_0\eta, \theta\rangle\,.
\end{equation}
The Majorana inner product that we use is\footnote{It is known that on
even-dimensional manifolds there are two Spin invariant Majorana inner
products. The other possibility, based on $C=i\Gamma_{03}$, was used
in \cite{Grover:2006wy}.}
\begin{equation}
A(\eta, \theta) = \langle C\eta^{\ast}, \theta\rangle\,, \label{Majorana}
\end{equation}
with the charge conjugation matrix $C=\Gamma_{12}$. Using the identities
\begin{equation}
\Gamma_a^{\ast} = -C\Gamma_0\Gamma_a\Gamma_0 C^{-1}\,, \qquad
\Gamma_a^T = -C\Gamma_a C^{-1}\,,
\end{equation}
it is easy to show that \eqref{Majorana} is Spin(3,1) invariant as well.

The charge conjugation matrix $C$ acts on the basis elements as
\begin{equation}
C 1 = e_{12}\,, \quad C e_{12} = -1\,, \quad C e_1 = -e_2\,, \quad C e_2 = e_1\,.
\end{equation}

In many applications it is convenient to use a basis in which the
gamma matrices act like creation and annihilation operators, given
by \eqn
\Gamma_{+}\eta\equiv\frac1{\sqrt2}\left(\Gamma_{2}+\Gamma_{0}\right)\eta
&=&\sqrt2\,e_{2}\rfloor\eta\,, \qquad
\Gamma_{-}\eta\equiv\frac1{\sqrt2}\left(\Gamma_{2}-\Gamma_{0}\right)\eta
=\sqrt2\,e_{2}\wedge\eta\,, \nonumber \\
\Gamma_{\bullet}\eta\equiv\frac1{\sqrt2}\left(\Gamma_{1}-i\Gamma_{3}\right)\eta
&=&\sqrt2\,e_{1}\wedge\eta\,, \qquad
\Gamma_{\bar\bullet}\eta\equiv\frac1{\sqrt2}\left(\Gamma_{1}+i\Gamma_{3}\right)\eta
=\sqrt2\,e_{1}\rfloor\eta\,.
\feqn
The Clifford algebra relations in this  basis are $\{\Gamma_A,\Gamma_B\} = 2\eta_{AB}$,
where $A,B,\ldots = +,-,\bullet,\bar\bullet$ and the nonvanishing components of
the tangent space metric read
$\eta_{+-} = \eta_{-+} = \eta_{\bullet\bar\bullet} = \eta_{\bar\bullet\bullet} = 1$.
The spinor 1 is a Clifford vacuum, $\Gamma_{+}1 = \Gamma_{\bar\bullet}1 = 0$,
and the representation $\Delta_c$ can be constructed by acting on 1 with the
creation operators $\Gamma^+ = \Gamma_-, \Gamma^{\bar\bullet} = \Gamma_{\bullet}$,
so that any spinor can be written as
\begin{displaymath}
\eta = \sum_{k=0}^2 \frac 1{k!}\phi_{{\bar a}_1\ldots {\bar a}_k}\Gamma^{{\bar a}_1
\ldots {\bar a}_k}1\,, \qquad \bar a = +,\bar\bullet\,.
\end{displaymath}
The action of the Gamma matrices and the Lorentz generators $\Gamma_{AB}$ is
summarized in table \ref{tab:gamma}.

\begin{table}[ht]
\begin{center}
\begin{tabular}{|c||c|c|c|c|}
\hline
& 1 & $e_{1}$ & $e_{2}$ & $e_{1}\wedge e_{2}$\\
\hline\hline
$\Gamma_{+}$ & 0 & 0 & $\sqrt2$ & $-\sqrt2e_{1}$\\
\hline
$\Gamma_{-}$ & $\sqrt2e_{2}$ & $-\sqrt2e_{1}\wedge e_{2}$ & 0 & 0\\
\hline
$\Gamma_{\bullet}$ & $\sqrt2e_{1}$ & $0$ & $\sqrt2e_{1}\wedge e_{2}$ & 0\\
\hline
$\Gamma_{\bar\bullet}$ & 0 & $\sqrt2$ & 0 & $\sqrt2e_{2}$\\
\hline\hline
$\Gamma_{+-}$ & 1 & $e_{1}$ & $-e_{2}$ & $-e_{1}\wedge e_{2}$\\
\hline
$\Gamma_{\bar\bullet\bullet}$ & 1 & $-e_{1}$ & $e_{2}$ & $-e_{1}\wedge e_{2}$\\
\hline
$\Gamma_{+\bullet}$ & 0 & 0 & $-2e_{1}$ & 0\\
\hline
$\Gamma_{+\bar\bullet}$ & 0 & 0 & 0 & 2\\
\hline
$\Gamma_{-\bullet}$ & $-2e_{1}\wedge e_{2}$ & 0 & 0 & 0\\
\hline
$\Gamma_{-\bar\bullet}$ & 0 & $2e_{2}$ & 0 & 0\\
\hline
\end{tabular}
\end{center}
\caption{The action of the Gamma matrices and the Lorentz generators
  $\Gamma_{AB}$ on the different basis elements. \label{tab:gamma}}
\end{table}

Note that $\Gamma_A = {U_A}^a\Gamma_a$, with
\begin{displaymath}
\left({U_A}^a\right) = \frac1{\sqrt2} \left(\begin{array}{cccc} 1 & 0 & 1 & 0 \\
-1 & 0 & 1 & 0 \\ 0 & 1 & 0 & -i \\ 0 & 1 & 0 & i
\end{array}\right) \in {\mathrm U}(4)\,,
\end{displaymath}
so that the new tetrad is given by $E^A = {(U^{\ast})^A}_a E^a$.

\end{document}